\documentclass[a4paper,12pt]{article}
\pdfoutput=1 

\usepackage{myjinst}
\usepackage{color}

\newcommand{\g}{g_{a\gamma \gamma }}
\newcommand{\red}{\protect\color{red}}
\newcommand{\blue}{\protect\color{blue}}
\newcommand{\black}{\protect\color{black}}

\renewcommand\blue\red
\renewcommand\red\black

\title{
Towards a medium-scale axion helioscope and haloscope}

\author[a]{V.~Anastassopoulos,}
\author[b]{F.~Avignone,}
\author[c]{A.~Bykov,}
\author[d]{G.~Cantatore,}
\author[e]{S.A.~Cetin,}
\author[f]{A.~Derbin,}
\author[f]{I.~Drachnev,}
\author[g]{R.~Djilkibaev,}
\author[c]{V.~Eremin,}
\author[h]{H.~Fischer,}
\author[i]{A.~Gangapshev,}
\author[a]{A.~Gardikiotis,}
\author[g]{S.~Gninenko,}
\author[g]{N.~Golubev,}
\author[j]{D.H.H.~Hoffmann,}
\author[k]{M.~Karuza,}
\author[g]{L.~Kravchuk,}
\author[g]{M.~Libanov,}
\author[l]{A.~Lutovinov,}
\author[a]{M.~Maroudas,}
\author[g,m]{V.~Matveev,}
\author[l]{S.~Molkov,}
\author[f]{V.~Muratova,}
\author[g]{V.~Pantuev,}
\author[l]{M.~Pavlinsky,}
\author[g]{K.~Ptitsyna,}
\author[g]{G.~Rubtsov,}
\author[f]{D.~Semenov,}
\author[n]{P.~Sikivie,}
\author[o]{A.~Spiridonov,}
\author[p]{P.~Tinyakov,}
\author[g]{I.~Tkachev,}
\author[g]{S.~Troitsky,}
\author[f]{E.~Unzhakov,}
\author[a]{and K.~Zioutas}

\affiliation[a]{Patras University, Patras, Greece}
\affiliation[b]{University of South Carolina, Columbia, USA}
\affiliation[c]{Ioffe Institute RAS, St.~Petersburg, Russia}
\affiliation[d]{University of Trieste, Italy}
\affiliation[e]{High Energy Physics Research Center, Bilgi University, Istanbul, Turkey}
\affiliation[f]{Petersburg Nuclear Physics Institute, St.~Petersburg,
Russia}
\affiliation[g]{Institute for Nuclear Research  RAS, Moscow,
Russia
}
\affiliation[h]{Albert-Ludwigs-Universitaet Freiburg, Germany}
\affiliation[i]{Baksan Neutrino Observatory, INR RAS, Neutrino, Russia}
\affiliation[j]{Institut f\"ur Kernphysik/Technische Universit\"at
Darmstadt, Germany}
\affiliation[k]{University Rijeka, Croatia}
\affiliation[l]{Space Research Institute  RAS, Moscow, Russia}
\affiliation[m]{Joint Institute for Nuclear Research, Dubna, Russia}
\affiliation[n]{University of Florida, Gainesville, USA}
\affiliation[o]{Physics Department, Moscow State University, Moscow,
Russia}
\affiliation[p]{Universit\'e Libre de Bruxelles, Brussels, Belgium}

\emailAdd{st@ms2.inr.ac.ru}

\abstract{
We discuss the physics case for and the concept of a medium-scale axion
helioscope with sensitivities in the axion-photon coupling a few times
better than CERN Axion Solar Telescope (CAST). Search for an axion-like
particle with these couplings is motivated by several persistent
astrophysical anomalies. We present early conceptual design, existing
infrastructure, projected sensitivity and timeline of
such a
helioscope (Troitsk Axion Solar Telescope Experiment, TASTE) to be
constructed in the Institute for Nuclear Research, Troitsk, Russia. The
proposed instrument may be also used for the search of dark-matter halo
axions.}


\begin{document}
\maketitle
\flushbottom

\section{Introduction and motivation}
\label{sec:motiv}
\subsection{Axions and axion-like particles}
\label{sec:motiv:axions}
The Standard Model of particle physics cannot explain the mechanism behind
the charge-parity (CP) symmetry conservation in strong interactions and
does not possess a viable candidate for a dark-matter particle (see e.g.\
Ref.~\cite{ST-UFN} for a brief review of these major problems and further
references). Long ago, attempts to attack the first problem led to the
concept of the \textit{axion}, a pseudoscalar particle which appears as a
necessary ingredient of the Peccei-Quinn solution \cite{PQ, Weinberg,
Wilczek} to the strong CP problem, which remains the most successful and
popular approach to the puzzle. Subsequently, it has been understood that
the axion is a prospective candidate for a dark-matter
particle~\cite{AxDM1, AxDM2, AxDM3}. Recent results of the Large Hadron
Collider (LHC), which exclude a large number of other popular dark-matter
candidates, make the axion dark matter even more plausible and revive
theoretical and experimental interest to light pseudoscalars.

A characteristic property of the axion, which opens a number of
possibilities to its laboratory and astrophysical searches, is its
two-photon coupling in the Lagrangean,
\begin{equation}
-\frac{1}{4} \g \phi F_{\mu \nu } \tilde{F}^{\mu \nu },
\label{*}
\end{equation}
where $\phi$ is the pseudoscalar field, $F_{\mu \nu }$ is the
electromagnetic field stress tensor and $\tilde F^{\mu \nu }=\frac{1}{2}
\epsilon^{\mu \nu \rho \lambda } F_{\rho \lambda }$ is its dual tensor. In
most axion models related to Quantum Chromodynamics (QCD), the coupling
(\ref{*}) arises from the axion mixing with the $\pi$ meson and is related
to the axion mass $m_{a}$ as~\cite{Kaplan, Srednicki}
\begin{equation}
\g =10^{-10}~\mbox{GeV}^{-1}\, C_{\gamma }\, \frac{m_{a}}{0.5~{\rm eV}},
\label{**}
\end{equation}
where the model-dependent coefficient $C_{\gamma }\sim 1$
(see, however, Refs.~\cite{Rubakov, Zurab, Maurizio2}). At the same time,
pseudoscalars with a similar interaction (\ref{*}) arise in numerous
extensions of the Standard Model of particle physics, see e.g.\
Ref.~\cite{Ringwald-rev} for a review. Depending on the model, they may or
may not respect Eq.~(\ref{**}) and may or may not solve the strong CP
problem. For some values of couplings they also may be dark-matter
candidates (see Ref.~\cite{ALP-DM} for the parameter space relevant for
the case when no other interactions beyond Eq.~(\ref{*}) are present). In
any case, however, thanks to the interaction (\ref{*}), these
\textit{axion-like particles} (ALPs) should manifest themselves in hot
dense plasmas where thermonuclear reactions take place as well as in
numerous other laboratory and astrophysical environments.

\subsection{Astrophysical constraints and indications}
\label{sec:motiv:astro}
Recent progress in astrophysics makes it possible to use observational
data to search for axions and ALPs, or to constrain their parameters.
Detailed studies of stellar energy losses constrain $\g \lesssim 6.6
\times 10^{-11}$~GeV$^{-1}$ at the {\red 95}\% confidence level
\cite{HBstars}, at the same time giving a weak indication in favour of the
presence of an axion or ALP with $\g \sim \left(2.9 \pm 1.8 \right) \times
10^{-11}$~GeV$^{-1}$ at the 68\% confidence level, see
Ref.~\cite{Maurizio} for a wider discussion. However, a much stronger
evidence for the existence of an ALP with the photon coupling in this
domain comes from the gamma-ray astronomy (for a brief review, see
Ref.~\cite{ST-mini}).

Indeed, the Universe is filled with background radiation, on which
energetic gamma rays produce electron-positron pairs \cite{Nikishov}. This
process limits the mean free path of energetic photons to a small fraction
of the Universe, strongly dependent on the photon energy. While a
comparison of gamma-ray spectra of blazars observed at various distances
with FERMI-LAT shows the presence of a certain distance-dependent flux
suppresion \cite{FERMI-suppression}, analyses of ensembles of gamma-ray
sources at distances corresponding to large optical depths indicate
\cite{HornsMeyer, RT} that the suppression is much weaker than expected.
The statistical significance of this anomaly, 12.4 standard
deviations~\cite{RT}, makes it a strong argument in favour of existence of
unaccounted processes related to the gamma-ray propagation. Interestingly,
all the studies which indicate the anomaly have been based on minimal
models of the extragalactic background light (EBL), e.g.\ \cite{Gilmore,
Franceschini}, on the level of the sum of the light from observed
galaxies~\cite{EBLlower-bounds}, while the very recent dedicated
observations making use of two different approaches \cite{newEBLobs1,
newEBLobs2}, indicate the EBL intensity twice higher. Proved to be true,
these EBL values would make the gamma-ray propagation anomaly even more
dramatic.

Potential astrophysical explanations in terms of secondary particles
\cite{Kusenko, Dzhatdoev} have troubles explaining the effect for most
distant sources and, more importantly, are at odds with the observations
of strong variability of gamma-ray sources at large optical depths, e.g.\
\cite{1501.05087}. One is forced to invoke new physics for the solution to
the anomaly. The pair-production probability might be modified in the
presence of a weak Lorentz-invariance violation; however, this violation
would also result in non-observation of any TeV photons by Cerenkov
atmospheric telescopes because the development of photon-induced air
showers would also be suppressed, and is therefore excluded \cite{noLIV1,
noLIV2}.

The remaining viable explanation points to ALPs. Thanks to the interaction
(\ref{*}), photon and ALP mix in external magnetic fields ~\cite{sikivie,
raffelt}, while the ALP does not produce $e^{+}e^{-}$ pairs. Depending on
the parameters, this may result either in axion-photon oscillations in
intergalactic magnetic fields, which would enlarge the mean free path of
photons from distant sources~\cite{Csaba, DARMA}, or in a conversion of a
part of emitted photons to ALPs in the magnetic field in the source or in
its close environment, subsequent propagation of these ALPs through the
Universe and reconversion back to photons in the Milky Way or its
surroundings~\cite{Serpico, FRT}. Present upper limits on extragalactic
magnetic fields together with constraints on ALP parameters from
non-observation of gamma radiation from supernova SN1987A, persistence of
the gamma-ray propagation anomaly up to high redshifts and some hints on
the Galactic anisotropy in the anomaly manifestation make the second
scenario more favourable \cite{gal-egal}, though the first one is not yet
excluded. The second, Galactic-conversion, scenario may be realized for
$\g \sim \left(10^{-11} - 10^{-10}  \right)$~GeV$^{-1}$ and $m_{a} \sim
\left( 10^{-9} - 10^{-7} \right)$~eV. Experimental searches for a particle
with parameters in this range is therefore strongly motivated.

\subsection{Landscape of experimental projects}
\label{sec:motiv:experiments}
There are three general approaches to the experimental searches of axions
and ALPs.
\subsubsection{Purely laboratory experiments}
\label{sec:motiv:experiments:LSW}
An interesting class of experiments is called ``Light Shining through
Walls''. In experiments of this kind (see e.g.\ Ref.~\cite{LSW} for a
review), a laboratory produced laser beam passes through the magnetic
field, where a small fraction of photons convert to axions or ALPs. These
particles penetrate freely through a nontransparent wall. Another region
of the magnetic field provides for conditions for the reconversion of
photons from ALPs. These photons, if any, are registered by a very
sensitive detector. In the laboratory conditions, the probability of
conversion for one pass of the photon through the installation is
negligible, therefore an optical cavity is installed in the conversion
magnet. This increases the probability of conversion proportionally to the
cavity {\red finesse}. The most precise experiments of this kind were ALPS
and OSQAR \cite{ALPS, OSQAR}. It has been shown theoretically
\cite{ResRegen}, that a second cavity, if installed in the reconversion
magnet and locked with the first one, gives a further increase in the
overall detection rate. This approach, called resonant regeneration, has
never been realized in practice, but is planned to be implemented in
ALPS-II experiment \cite{ALPS-II}.  Another purely laboratory approach to
the axion/ALP searches is based on polarization measurements, in which the
vacuum birefringence in the external magnetic field is searched for. The
most precise experiments of this kind was PVLAS \cite{PVLAS}.

\subsubsection{Axion dark matter searches}
\label{sec:motiv:experiments:DM}
Axions constituting the Milky Way dark matter halo resonantly convert into
(almost) monochromatic microwave signal $\omega = m_a$ in a high-Q
microwave cavity permeated by a strong magnetic field
\cite{sikivie}. Such axion search experiments  are called {\it
haloscopes}. The Axion Dark Matter eXperiment (ADMX) has been pursuing
this technique since 1996 and ruled out a range of axion models with 1.9~$\mu$eV
$< m_a < 3.69~\mu$eV \cite{Asztalos:2009yp}. Recently ADMX-HF
\cite{Brubaker:2016ktl} has put strong limits $g_{a\gamma\gamma} <  2
\times 10^{-14}~{\rm GeV}^{-1}$ over the mass range 23.55~$\mu$eV $< m_a <
24.0 ~\mu$eV.
\red These experiments are probing parameter ranges predicted by realistic
axion models where axions can constitute a sizeable fraction of dark
matter in the Universe. \black Corresponding limits (together with early
experimental results) are shown in Fig.~\ref{fig:m-g} by green shaded
areas.

Technologies enabling dark matter detection at higher axion masses are
urgently needed, in particular, probing the interesting and
phenomenologically reach class of models, where the Peccei-Quinn phase
transition occurs after inflation. Such experiment{\red s}  ha{\red ve} been
proposed recently \red \cite{Majorovits:2016yvk,
MADMAXinterestGroup:2017bgn, Rybka} \black for the search of dark matter
axions in the mass range (40--400)~$\mu$eV. \red  They use novel detector
architecture which can be referred to as Open Cavity Resonators. \black The
experiment \red \cite{Majorovits:2016yvk,
MADMAXinterestGroup:2017bgn}\black, called MADMAX, will consist of 80
semi-transparent parallel disks with area $\sim$1~m$^2$ made from material
with high dielectric constant and placed in a strong magnetic field of 10
Tesla. Separations of the discs can be adjusted which would allow to probe
the emission of axion induced electromagnetic waves in the 10-100 GHz
domain, with the frequency given by the axion mass. It is planned that  as
a first step a smaller prototype with disc diameters of $\sim$30 cm will
be designed and produced in the next 2-3 years.

\red The experiment \cite{Rybka}, called Orpheus, will use open Fabry-Perot
resonator and a series of current-carrying wire planes. Orpheus design
assumes smaller magnetic field, 3 or 6 Tesla, depending upon targeted
axion mass range.

The axion dark matter mass window will be probed in the upcoming decade
also by other axion dark matter direct detection experiments as well. In
addition to MADMAX and Orpheus, the projects include CULTASK
\cite{CULTASK}, HAYSTAC \cite{Brubaker:2017rna}, CASPEr
\cite{Garcon:2017ixh}, for a recent reviews see \cite{Graham:2015ouw,
Petrakou:2017epq}. Some concepts of direct axion dark matter searches can
be implemented in helioscopes, as it was done recently in  CAST
\cite{IBS}. Most interesting for us is the MADMAX concept (or
dish antenna \cite{Horns:2012jf}) since it is relatively broadband and
allows for searches of amplified rare signals from streaming dark matter
\cite{Tinyakov:2015cgg, Zioutas:2017klh}, see Section 7.2.1. \black

\subsubsection{Search for solar axions}
\label{sec:motiv:experiments:solar}
Like other stars, our Sun contains a huge thermonuclear reactor in its
center, and axions or ALPs, if exist, should be produced there. They can
be detected on the Earth with an axion helioscope \cite{sikivie}, a
tube pointing to the Sun and filled with magnetic field allowing for
ALP-photon conversion and subsequent photon detection. Since the
helioscope is the concept we choose for our proposal, we discuss it in
more detail in Sec.~\ref{sec:helio}, \ref{sec:fom}. The CERN Axion Solar
Telescope (CAST) is, up to day, the most powerful helioscope which has
recently delivered the world-best upper limit on $\g$ \cite{CAST2017}.
Amusingly, a weak excess of events was found in some runs, but it is not
statistically significant. CAST has now finished its solar axion runs. An
ambitious new project, the International Axion Observatory (IAXO), has
been proposed a few years ago \cite{NGAH, IAXO-CDR} and is now at the
research and design stage. Another possibility to search for solar axions
on the Earth is to explore their hadronic couplings, see e.g.\
Refs.~\cite{Debrin, Gangapshev}.

\subsubsection{Limits and sensitivities}
\label{sec:motiv:experiments:limits}
\begin{figure}
\center{\includegraphics[width=1\linewidth]{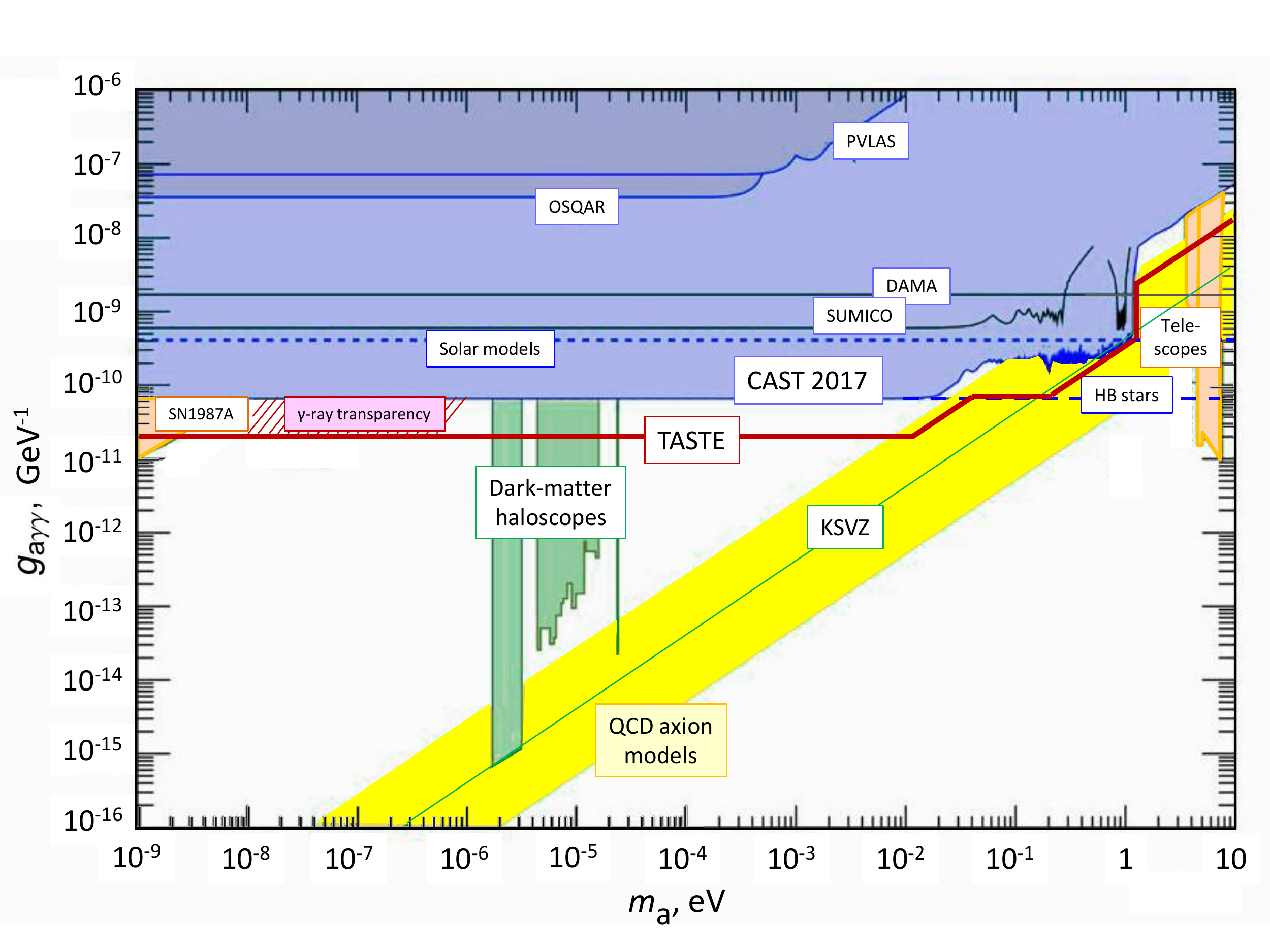}}
\caption{
The axion/ALP parameter space $(m_{a},\g)$.
Shaded areas are excluded by light-shining-through-walls experiment
OSQAR~\cite{OSQAR}, polarization experiment PVLAS~\cite{PVLAS2016}, solar
axion searches by helioscopes Sumico~\cite{SUMICO} and
CAST~\cite{CAST2017}, as well as by the DAMA experiment~\cite{DAMAaxions},
by telescope non-observation of cosmic axion decay
lines~\cite{telescope-limits1, telescope-limits2, telescope-limits3}  and
by non-observation of gamma radiation from supernova
1987A~\cite{SN1987Agamma}. The haloscope excluded areas
\cite{Asztalos:2009yp, Brubaker:2016ktl, halo1, halo2} assume that axions
constitute the galactic dark matter. Horizontal dashed lines give
astrophysical upper limits on $\g$ from solar data \cite{solar-models} and
from energy losses of horizontal-branch (HB) stars~\cite{HBstars}. The
yellow band indicates the parameter range favoured by QCD axion models,
with the green KSVZ line corresponding to a benchmark example \cite{K,
SVZ}. }
\label{fig:m-g}
\end{figure}

Presently, none of the experiments has demonstrated an evidence for an
axion or ALP. Astrophysical and experimental limits on the $(m_{a},\g)$
parameter space are shown in Fig.~\ref{fig:m-g},
where we also indicate by hatching the range of parameters favoured by the
Galactic photon/ALP mixing explanation of the gamma-ray absorption
anomaly. While the best limits on $\g$ in the relevant mass range,
$\g<6.6\times 10^{-11}$~GeV$^{-1}$ (95\% CL), is provided by the CAST
final results, one can see that the astrophysically motivated range of
couplings is just a few times lower. Still, we will see in
Sec.~\ref{sec:fom} that to reach this level of sensitivity, one needs an
instrument with the signal-to-noise ratio $\sim 100$ times better than
CAST. In this paper, we propose to build a new helioscope with the
sensitivity down to $\g\approx 2\times 10^{-11}$~GeV$^{-1}$ for a wide
range of $m_{a}$, covering the region of the parameter space motivated by
astrophysics, with the possibility to be used as an axion dark-matter
haloscope at the same time.

\begin{figure}
\center{%
\includegraphics[width=1\linewidth]{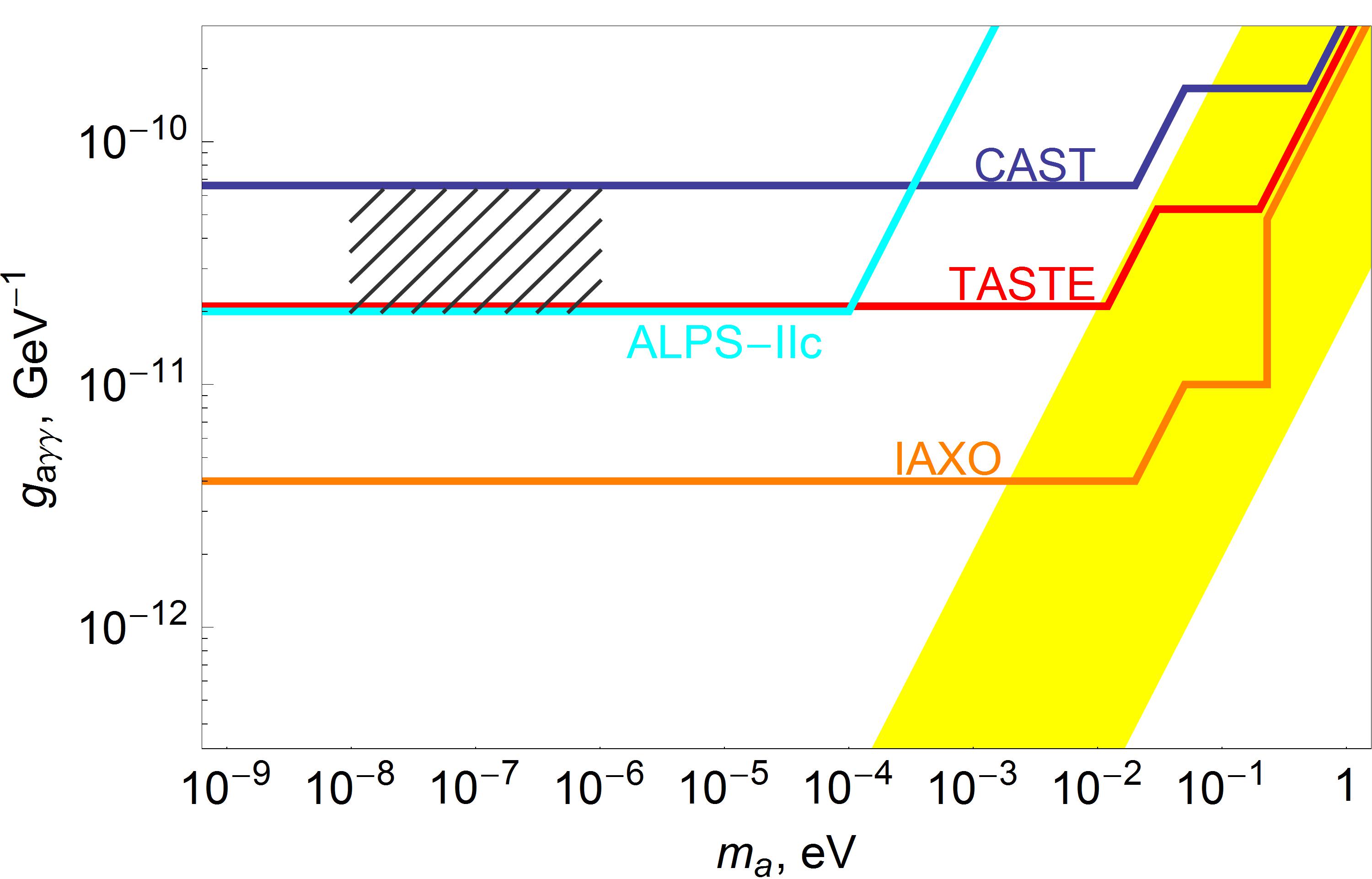}
}
\caption{
A sketch of comparison of sensitivities of proposed experiments to ALP
parameters with the CAST limits. The yellow band is favoured by QCD axion
models, the hatched area is favoured by the Galactic ALP conversion
scenario explaining the anomalous transparency of the Universe.}
\label{fig:sens}
\end{figure}

In Fig.~\ref{fig:sens}, we compare the projected sensitivity of our
helioscope (Troitsk Axion Solar Telescope Experiment, TASTE) with those of
two other projects aimed to explore the axion-photon coupling beyond the
CAST limits, IAXO and ALPS-IIc.
Both projected experiments plan to cover the range of the parameter space
motivated by the gamma-ray transparency of the Universe. However, there
are significant differences with our proposal, which make all three
projects complementary.

Indeed, ALPS-IIc, a light-shining-through-wall experiment, will be based
on the resonant-regeneration technique, which has not been demonstrated at
work yet. If it works as planned, the first scientific runs are expected
in 2020. Compared to helioscopes, this experiment will not cover the
region of higher-mass ALPs and therefore will not explore the standard QCD
axion scenario.

IAXO is a huge next-generation axion helioscopes with expected sensitivity
superceding other projects. Given its scale and cost, the start of the
full-scale experiment is planned beyond 2022. TASTE may be considered as
\red the first step towards IAXO, \black aimed to scan physically
interesting ALP and axion parameter space at much shorter timescale and at
much lower cost.

\section{The helioscope concept}
\label{sec:helio}
\begin{figure}
\center{\includegraphics[width=1\linewidth]{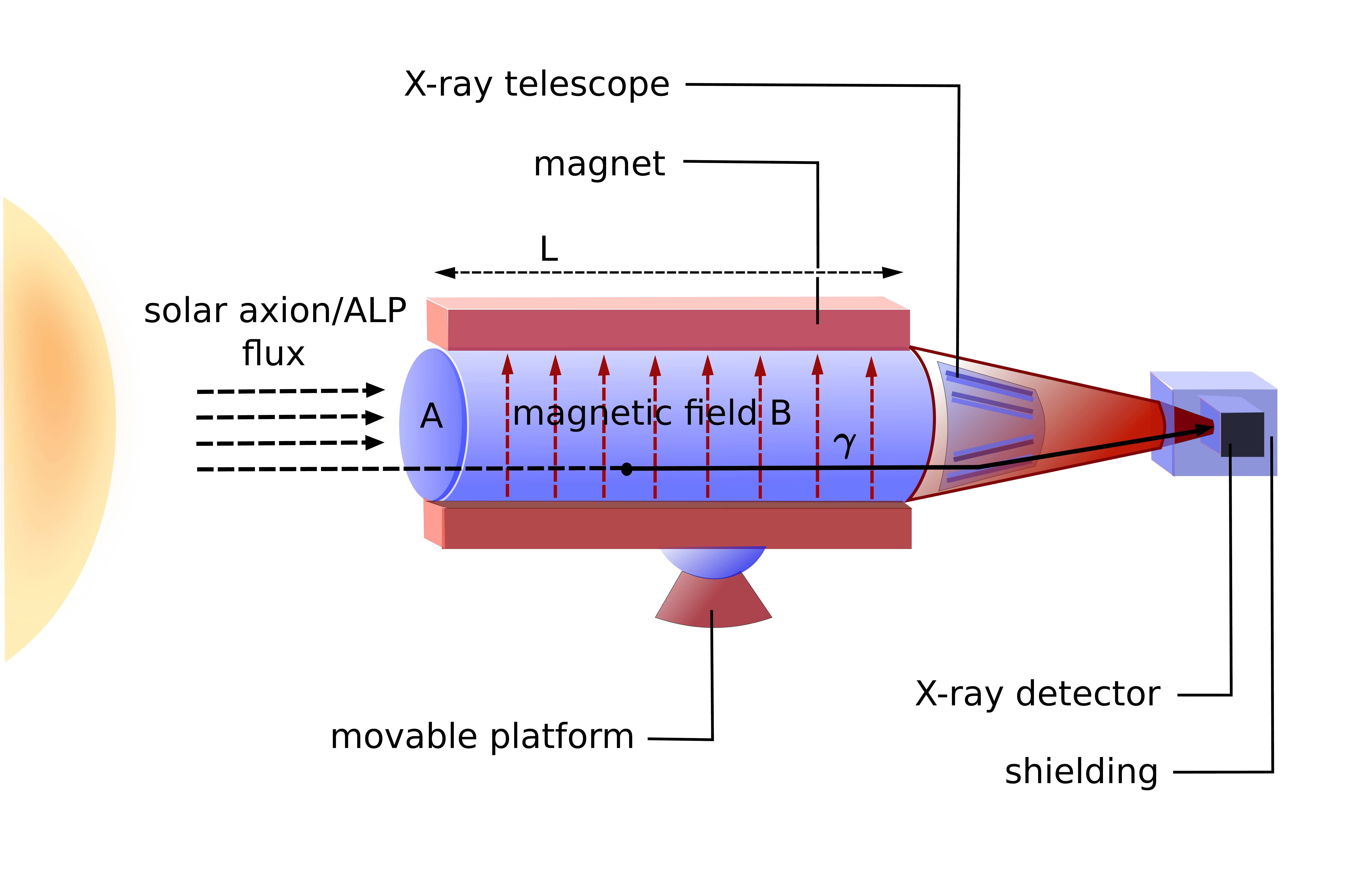}}
\caption{
The sketch of the helioscope concept.
}
\label{fig:helioscope}
\end{figure}

In this section, we briefly review the concept which we plan to apply to
searches for axions and ALPs, first proposed in Ref.~\cite{sikivie}. The
sketch of the approach is presented in Fig.~\ref{fig:helioscope}.
Axions and other light particles can be produced in hot and dense plasma in
the Sun's core; unlike photons, they can penetrate freely through its outer
parts and reach subsequently the terrestrial laboratory, entering the
magnet tube shielded from any light. A tiny fraction of the solar axions is
converted to photons in the field of the magnet and may be registered by a
sensitive detector. To reduce the background, these photons may be focused
into a small spot. Since the mean energy of the ALPs produced in the Sun
are determined by the solar core temperature (keVs), a modern helioscope
should include an X-ray telescope and a low-background detector of single
X-ray photons.

\subsection{Axion production in the Sun}
\label{sec:helio:Sun}
Axions and other light particles are produced in hot plasmas in the
regions where thermonuclear reactions take place, in particular in the
central parts of the Sun. For a general ALP, the guaranteed production
channel is the conversion of thermal photons to pseudoscalars
due to the interaction (\ref{*}) on the external field of the medium
nuclei and electrons\footnote{This conversion is often called the Primakoff
process, but, as suggested by one of us (KZ), a more proper name is
the Sikivie~\cite{sikivie} process: the Primakoff effect involves $\pi$
mesons and it was proposed~\cite{Primakoff} in 1951, decades before axions
have been introduced in physics.}. If the ALP has other interactions
besides Eq.~(\ref{*}), other channels may dominate, in particular those
related to the tree-level electron coupling. Since these production
channels are less general, we postpone their discussion to
Sec.~\ref{sec:concl:opport:electron} and concentrate on the $\g$-related
flux here.

\begin{figure}
\center{\includegraphics[width=0.85\linewidth]{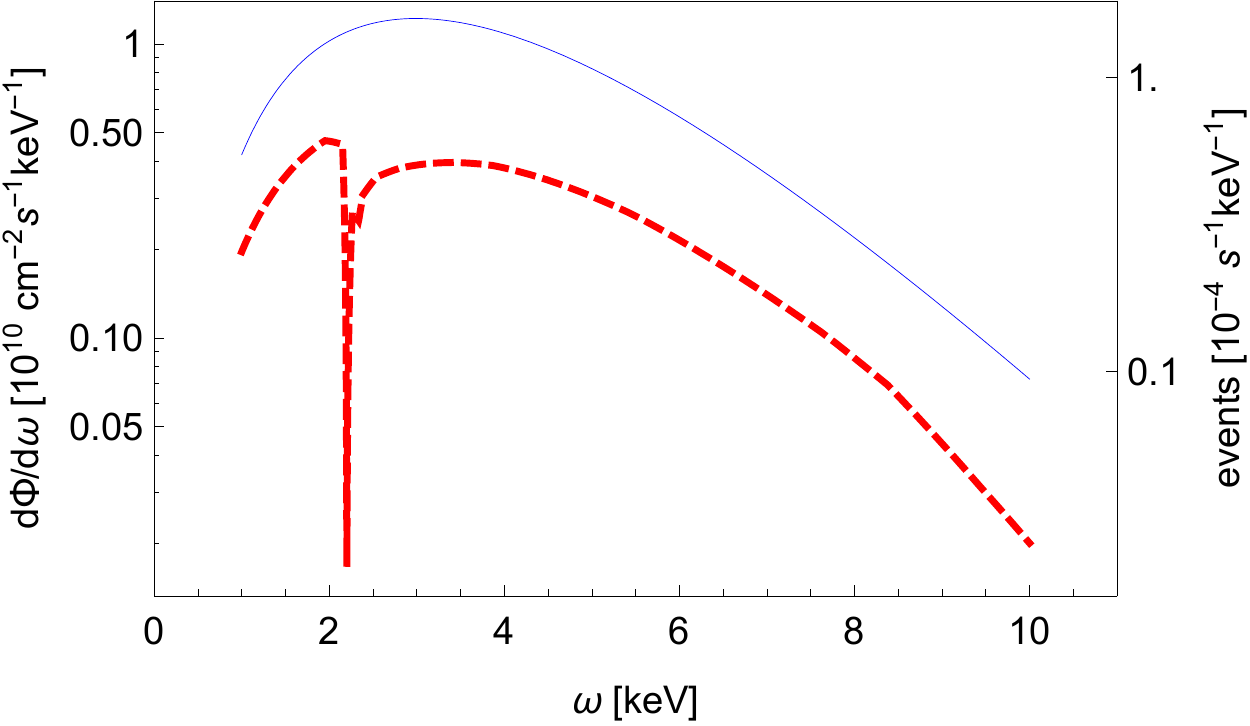}
\caption{
\red Full line: \black the flux \red at the Earth \black of solar axions or
ALPs emitted due to the Sikivie process~\cite{CAST2007} for $\g=4\times
10^{-11}$~GeV$^{-1}$ and $m_{a}\lesssim 0.01$~eV. The right-hand-side
scale gives the number of converted photons in TASTE per second, assuming
$B=3.5$~T, $L=12$~m and the tube diameter of 60~cm. Note that the flux
scales as $\g^{2}$ but the number of converted photons scales as $\g^{4}$.
\red The dashed line assumes SODART energy-dependent efficiency, see
Sec.~\ref{sec:X:opt}, and gives the number of photons collected by the
telescope.\black
\label{fig:flux}
}
}
\end{figure}

The flux of ALPs produced in the solar interior by means of the
interaction (\ref{*}) has been calculated e.g.\ in Ref.~\cite{CAST2007}
and is presented, for typical values of parameters, in Fig.~\ref{fig:flux}.
A convenient parametrization, valid for ALP energies 1~keV$\lesssim \omega
\lesssim 11$~keV, is given~\cite{CAST2007} by
\begin{equation}
\frac{d\Phi}{d\omega} = \mathcal{ N}
\left(\frac{\g}{10^{-10}~\mathrm{GeV}^{-1}} \right)^{2}
\left(\frac{\omega}{\omega_{0}} \right)^{\beta}
\exp \left(-\left(\beta +1 \right) \frac{\omega}{\omega_{0}} \right),
\label{Eq:flux-par}
\end{equation}
where the normalization constant is $\mathcal{ N}= 2.11\times
10^{12}$~cm$^{-2}\,$s$^{-1}\,$keV$^{-1}$, the mean energy is
$\omega_{0}=4.195$~keV and the parameter $\beta=2.481$. For a broad range
of ALP parameters, the flux scales as $\g^{2}$ and does not depend on
$m_{a}$.

\subsection{Conversion in the helioscope}
\label{sec:helio:Pconv}
Suppose that the helioscope magnetic field, transverse to the direction to
the Sun, is $B$, constant over the vacuum conversion zone of length $L$.
Then the probability to create a photon from an axion/ALP
is~\cite{sikivie, raffelt}
\begin{equation}
\displaystyle
P_{a\to \gamma }=
4.6 \times 10^{-18}
\left(\frac{\g}{10^{-10}~\mathrm{GeV}^{-1}} \right)^{2}
\left(\frac{B}{3.5~\mathrm{T}}  \right)^{2}
\left(\frac{L}{12~\mathrm{m}}  \right)^{2}
\frac{\sin^{2}(qL/2)}{(qL/2)^{2}},
\label{Eq:Pconv}
\end{equation}
where in the vacuum case, $q=m_{a}^{2}/(2\omega)$.
The last multiplier, sometimes called the form factor, is due to the
nonzero $m_{a}$ and describes the loss of coherence between the photon
and ALP wave functions; it is almost constant (coherent conversion)
provided
\begin{equation}
m_{a} \lesssim 0.012~\mathrm{eV}\,
\left(\frac{\omega}{\omega_{0}}\right)^{1/2}
\left(\frac{L}{12~\mathrm{m}}\right)^{-1/2}.
\label{Eq:11*}
\end{equation}
If the conve{\red r}sion zone is filled with a buffer gas, $m_{a}^{2}$ in
Eqs.~(\ref{Eq:Pconv}), (\ref{Eq:11*}) should be replaced with
$(m_{a}^{2}-m_{\gamma }^{2})$, where $m_{\gamma }$ is the effective photon
mass. The latter is, under the conditions we discuss, equal to the plasma
frequency, $m_{\gamma }=\omega_{\rm pl}=\sqrt{4\pi \alpha n_{e}/m_{e}}$,
where $n_{e}$ is the electron concentration and $m_{e}$ is the electron
mass, $\alpha $ is the fine-structure constant. Therefore, matching $m_{a}
\approx m_{\gamma }$ helps to restore coherence at larger $m_{a}$ with the
same $L$. In our project, we will preview the possibility to fill the
conversion zone with gas, which would allow to extend the sensitivity
deeper into the parameter region favoured by the QCD axion models.

\subsection{Comparison to other approaches}
\label{sec:helio:comparison}
Helioscopes, compared to other instruments, test a very wide range of
ALP masses with the same observation. The result relies on robust
assumptions about physical conditions in the solar interior. Axion
dark-matter experiments may reach better sensitivity in $\g$ but only for
a very narrow range of $m_{a}$; what is more important, they rely
theoretically on particular assumptions about the axion interactions and
on the very early Universe. Modern light-shining-through-walls
experiments, though having a benefit of a pure laboratory approach, are
considerably less sensitive, while more involved and more sensitive future
implementations are based on yet unexplored techniques. The helioscope
concept has been succesfully tested, since the first realization in 1992
in Brookhaven \cite{Brookhaven}, with the Tokyo helioscope (SUMICO)
\cite{SUMICO} and the CERN axion solar telescope (CAST) \cite{CAST2017,
CAST2007, CAST99}. The novelty of our project is \red mostly \black in its
scale \red (e.g.\ a factor of 100 enhancement of the collecting area with
respect to CAST) \black and in the \red potential \black use of \red new
dedicated \black low-background X-ray \red detectors. \black

\section{Figures of merit and technical requirements}
\label{sec:fom}
A useful set of figures of merit, allowing to compare sensitivities of
axion helioscopes, has been discussed in Ref.~\cite{NGAH}. In this
section, we use the figures of merit to determine basic technical
requirements for a helioscope sensitive to the astrophysically motivated
range of $\g$. \red Clearly, this approach does not capture many important
details like the energy and directional dependence of the signal,
efficiency and background; a more detailed simulation of the TASTE
sensitivity will be performed upon fixing its design. \black

In the regime when the \red statistics of background counts is high
enough, \black the instrument's sensitivity is roughly driven by the ratio
$s/\sqrt{b}$, where $s$ and $b$ are the numbers of signal and background
events, respectively. The number of signal photons produced in the
conversion zone per second is given by a product of the flux
(\ref{Eq:flux-par})  and the conversion probability (\ref{Eq:Pconv}). The
total number of signal events in the detector is given by the time
integral of this rate multiplied by the detector efficiency. Overall, we
have
\[
s \propto \g^{4}\, B^{2} L^{2} A \cdot \epsilon_{t} t \cdot \epsilon_{o}
\cdot \epsilon_{d},
\]
where $A$ is the conversion-zone cross section area, $\epsilon _{t}$ is
the fraction of time the helioscope tracks the Sun (determined, in
particular, by the moving abilities of the mount), $t$ is the total
operation time, $\epsilon _{o}$ and $\epsilon _{d}$ are the efficiencies
of X-ray \red optics \black and detect\red{}or\black, respectively.

The number of background events is assumed to scale with the area $a$ of
the spot to which the X-ray telescope phocuses the photons,
\[
b= \bar{b}a\cdot \epsilon _{t}t,
\]
where $\bar{b}$ is the background count rate per unit area of the
detector. The overall sensitivity is therefore
\[
\frac{s}{\sqrt{b}} \propto \g^{4}\, F,
\]
where the figure of merit $F$ is determined as
\begin{equation}
F= B^{2}L^{2}A \,\cdot\, \frac{\epsilon _{o}\epsilon
_{d}}{\sqrt{a\,\bar{b}}} \,\cdot\, \sqrt{\epsilon _{t}t},
\label{Eq:FoM}
\end{equation}
where the three multipliers correspond to the magnet, the X-ray part and
the tracking time, respectively.

Since the incoming solar-ALP flux is the same for all devices, it is
customary to determine the sensitivity of a new helioscope to $\g$ through
the CAST sensitivity,
\[
\frac{\g}{g_{\rm CAST}} \simeq \left(\frac{F}{F_{\rm CAST}} \right)^{1/4},
\]
where $g_{\rm CAST}=6.6 \times 10^{-11}$~GeV$^{-1}$ is the CAST limit and
$F_{\rm CAST}$ is the figure of merit (\ref{Eq:FoM}) calculated for the
CAST parameters (note that these estimates are valid for the
coherent-conversion case, $m_{a}\lesssim 0.01$~eV).
\red This is a simplified approximation because the CAST result comes from
a combined analysis of different runs performed under different
conditions; however, this approach fits the overall level of precision
of our approximate estimates. \black The relevant CAST parameters are given
in Table~\ref{tab:FoM}. Our goal is to have $F/F_{\rm CAST} \sim 3^{4}$,
and this determines technical requirements for the experiment.
\begin{table}
\centering
\caption{\label{tab:FoM}
Parameters of TASTE versus CAST and their contributions to the figure of
merit. } \smallskip
\begin{tabular}{cccc}
\hline
parameter&CAST value&TASTE value& $F/F_{\rm CAST}$\\
         &(2017) & (projected) & contribution\\
\hline
$B$, T & 9 & 3.5 & 0.15\\
$L$, m & 9.26 & 12 & 1.68\\
$A$, m$^{2}$ & 0.003 & 0.28 & 94.2\\
\hline
in total (magnet)& & & 23.9\\
\hline
$\epsilon _{o}$ & \red 0.27 & \red 0.36 & \red 1.33\\
$\epsilon _{d}$ & 0.7 & 0.7 & 1\\
$\bar{b}$, cm$^{-2}$s$^{-1}$keV$^{-1}$ & $10^{-6}$ & $5\times 10^{-7}$ &
1.41\\
$a$, cm$^{2}$ & 0.15 & 0.5 & 0.55\\
\hline
in total (X-ray) & & & \red 1.03\\
\hline
$\epsilon _{t}$ & 0.12 & 0.5 & 2.04\\
$t$, years & 1.08 & 3 & 1.67\\
\hline
in total (tracking) & & & 3.4\\
\hline
\hline
$F/F_{\rm CAST}$ & & & \red 83.9\\
\hline
\end{tabular}
\end{table}

The principal benefit of the new device with respect to CAST will be in
the cross section of the conversion zone. Indeed, CAST used a
decommissioned magnet from the Large Hadron Collider (LHC) with $A_{\rm
CAST}\simeq 3\times 10^{-3}$~m$^{2}$ (in two bores). For TASTE, we will
construct a dedicated magnet so we are free to enlarge the cross-section
area. However, the limitation comes from the X-ray telescope: the largest
available ones have the diameter of $\sim 60$~cm. We therefore keep this
diameter fixed in our proposal. On the other hand, as will be discussed in
Sec.~\ref{sec:magnet}, we plan to use available superconducting cable,
whose parameters and amount determine the working magnetic field value
$B\sim 3.5$~T and the magnet length $L\sim 12$~m. Our goal is to have the
tracking time fraction $\epsilon _{t}\sim 0.5$ (going beyond that is not
only technically unfeasible but also poorly motivated, since we need a
sufficient amount of off-Sun time to measure the background). For the
X-ray optics, we use parameters of the SODART telescope, see
Sec.~\ref{sec:X}, that is the
collection efficiency $\sim$\red0.36\black. The area of the image in the
focal plane $a=0.5$~cm$^{2}$ is determined from Eq.~(3.8) of
Ref.~\cite{IAXO-CDR}. For the detector parameters $\epsilon _{d}$ and
$\bar b$, we take the best available presently values. These parameters,
summarized also in Table~\ref{tab:FoM}, result in the figure of merit
improvement $F/F_{\rm CAST}\sim 3^{4}$, which implies the expected
sensitivity 3 times better than CAST, that is down to $\g \sim 2\times
10^{-11}$~GeV$^{-1}$, see Figs.~\ref{fig:m-g}, \ref{fig:sens}.
\red More details on feasibility of these values are given below. \black

\section{Magnet and cryogenics}
\label{sec:magnet}
Two principal complications drive the preliminary magnet design which we
present here. Firstly, the magnetic field should be perpendicular to the
tube axis. Secondly, the entire system should be installed on a moving
mount and hence its weight should be minimized. We therefore select a
dipole-like magnet with active (iron-free) shielding, inspired in
particular by some of proposals for the detector magnets of the Future
Circular Collider (FCC), see e.g.\ Ref.~\cite{FCC}. Active shielding
implies the use of additional external coils to close magnetic flux
lines and to suppress stray fields. For large magnets, this approach is
advantageous compared to the conventional iron-yoke design \cite{Tigra}.

\begin{figure}
\center{\includegraphics[width=\linewidth]{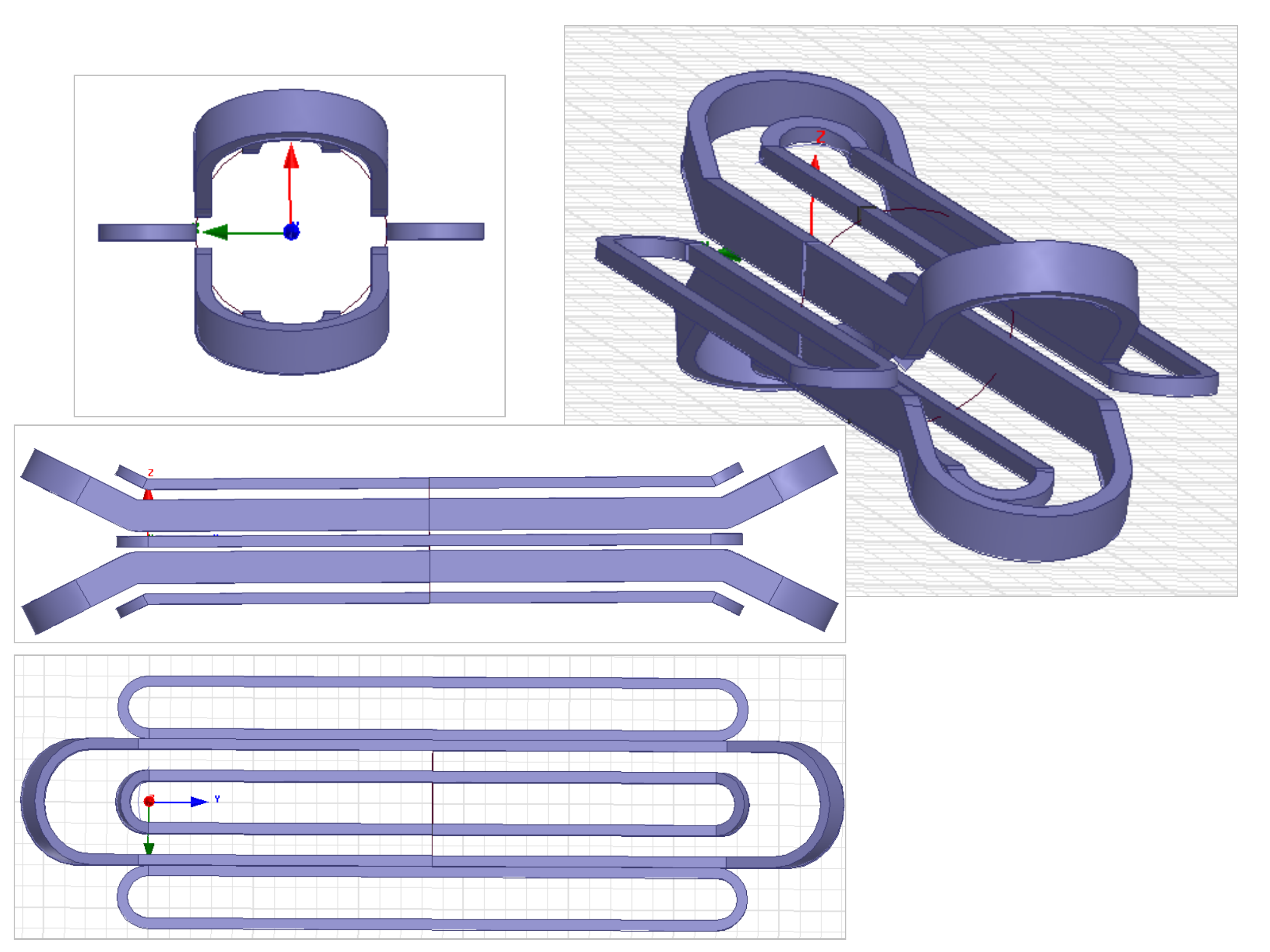}
\caption{Preliminary design of one section of the TASTE magnet: overview
of the coils and three projections. }
\label{fig:magnet-sec1}
}
\end{figure}

\begin{figure}
\center{\includegraphics[width=\linewidth]{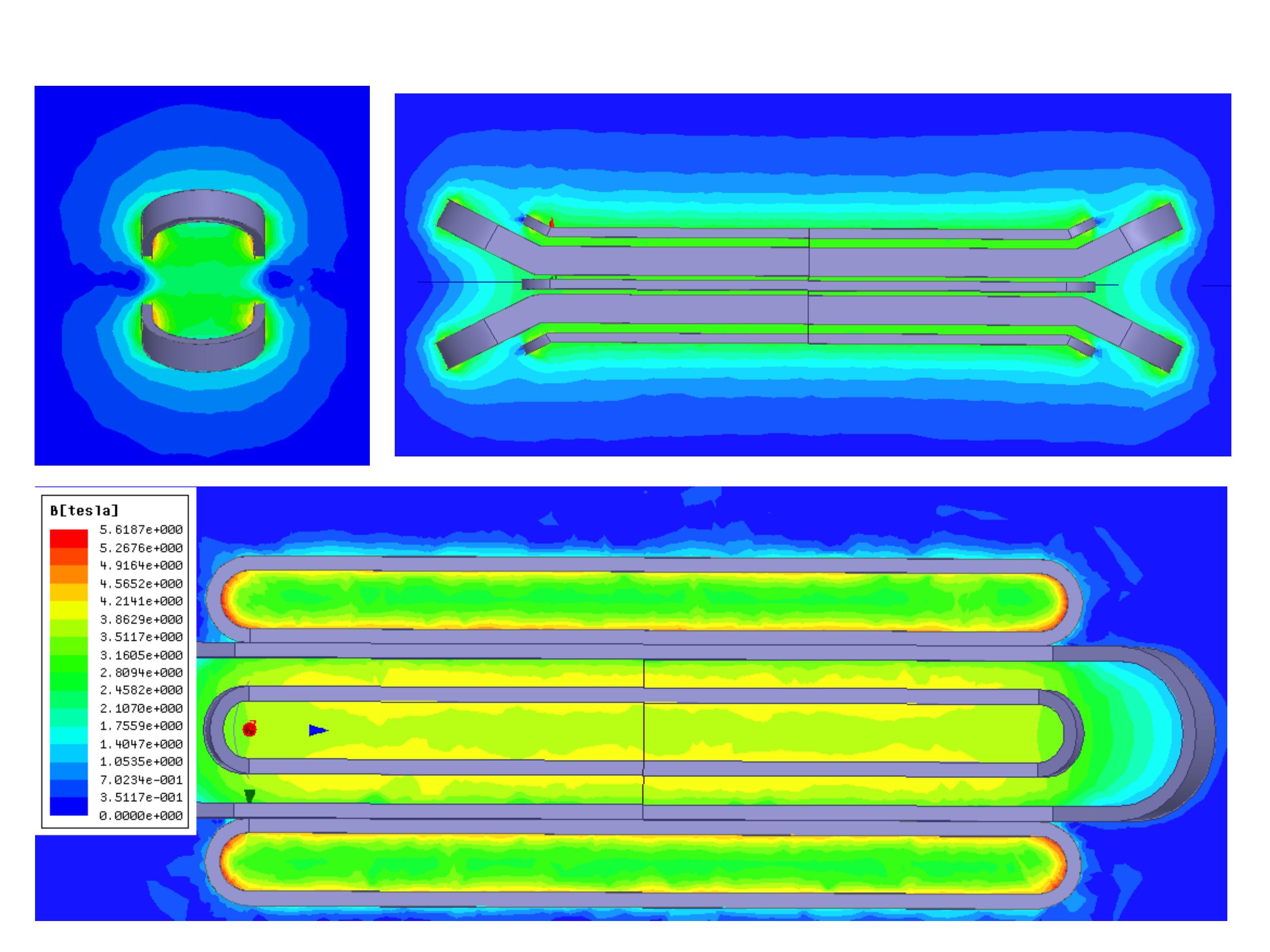}
\caption{Magnetic-field strength map for the design presented in
Fig.~\ref{fig:magnet-sec1}. }
\label{fig:magnet-sec2}
}
\end{figure}

The magnet, in our preliminary conceptual design, consists of three
identical sections, each of $\sim 4$~m length. The bore diameter is 60~cm,
as dictated by the X-ray telescope part. The bore will be kept cold in
order to possibly host equipment for dark-matter axion searches at certain
stages of the project, see Sec.~\ref{sec:concl:opport:DM}. The coil
configuration and the magnetic-field map for one section are presented in
Figs.~\ref{fig:magnet-sec1} and \ref{fig:magnet-sec2}, respectively.

Our plan is to construct one section first and to test it without the
moving mount and the X-ray telescope; we call this stage of the experiment
``LabTASTE'' because the 4-meter magnet with a cold bore may be used as a
laboratory to test various approaches to axion searches. It will be
sufficient to perform all dark-matter experiments described in
Sec.~\ref{sec:concl:opport:DM}. In parallel, depending on the availability
of funds, two other magnet sections will be manufactured and RnD works for
the X-ray part, as well as to the technical design of the moving platform,
will be finalized.

\begin{figure}
\center{%
\includegraphics[width=0.75\linewidth]{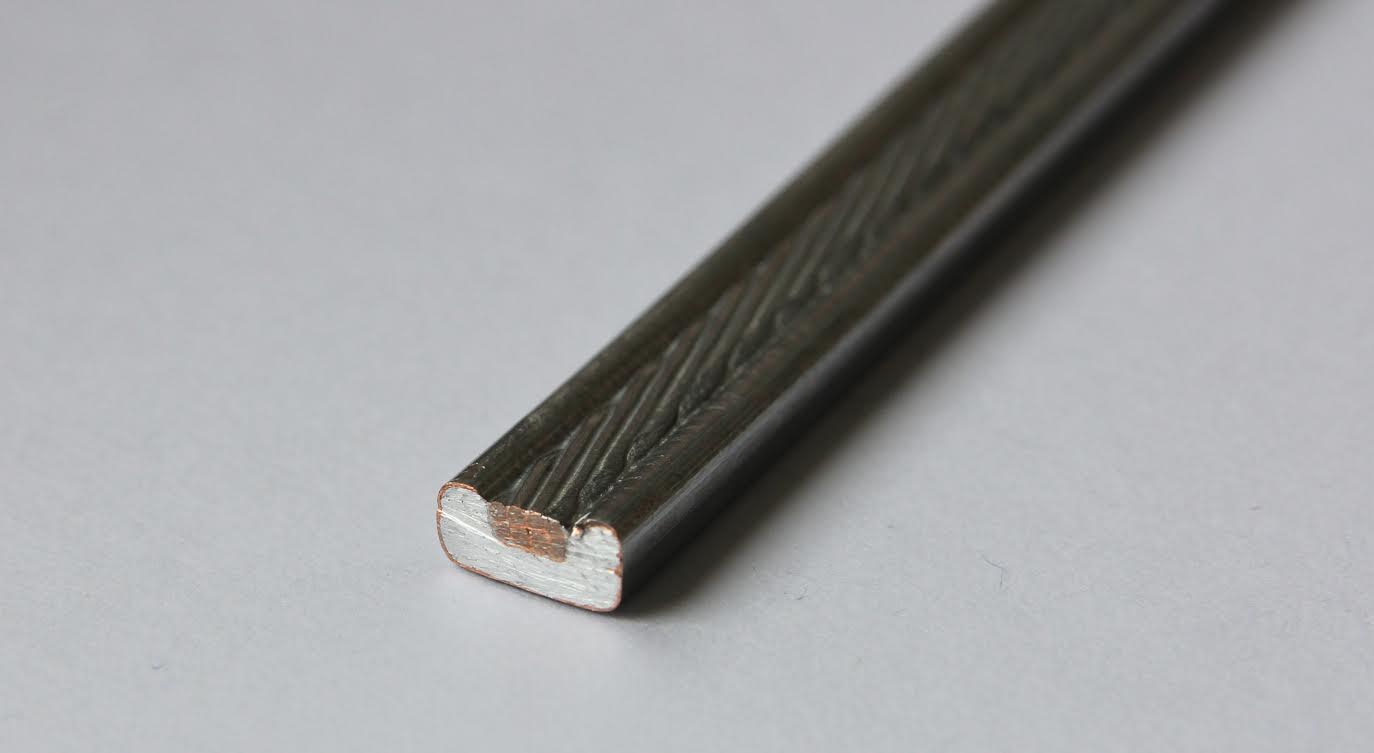}
\caption{The
superconducting cable of the MELC experiment, planned
to be used in TASTE.
}
\label{fig:wire}
}
\end{figure}

To make the magnet, we plan to use $\sim 35$~km of superconducting cable
available at INR. It has been manufactured in 1990s for the MELC
experiment~\cite{MELC} proposed to search for $\mu-e$ conversion in INR,
Troitsk. This experiment has never been launched but the conductor
(Fig.~\ref{fig:wire})
is still kept in INR. The cable consists of 8 twisted veins of the
PSNT-0.85 NiTi superconducting wire soldered in the aluminium-copper
matrix with the cross section of $4\times 9$~mm$^{2}$. The high-purity
aluminium (99.995\%) is placed into the copper matrix (the thickness of
the copper layer is $\approx 0.1$~mm). The critical current at 4.2~K was
measured in the outer magnetic field of 3~T to be $\approx (6.5-7.0)$~kA.
Measurements at different values of the field and theoretical estimates
suggest that the conductor can be safely used in magnetic fields of $\sim
5$~T at the current of $\sim 3.5$~kA, which is implied by our design. We
note that the magnet design presented here is very preliminary; parameters
of the magnet should be optimized at the technical-design stage.

\begin{figure}
\center{%
\includegraphics[width=0.492\linewidth]{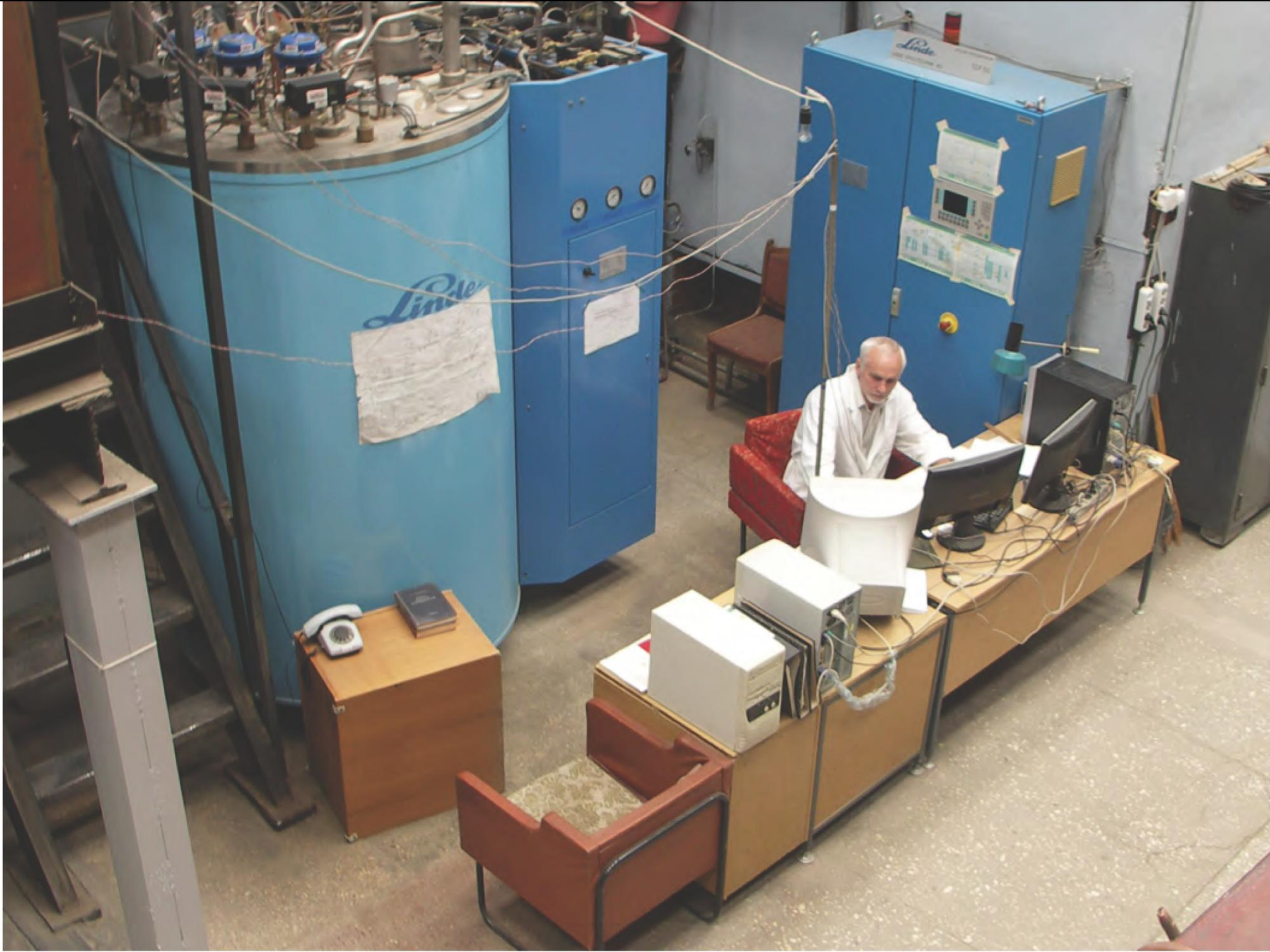}
\hfill
\includegraphics[width=0.492\linewidth]{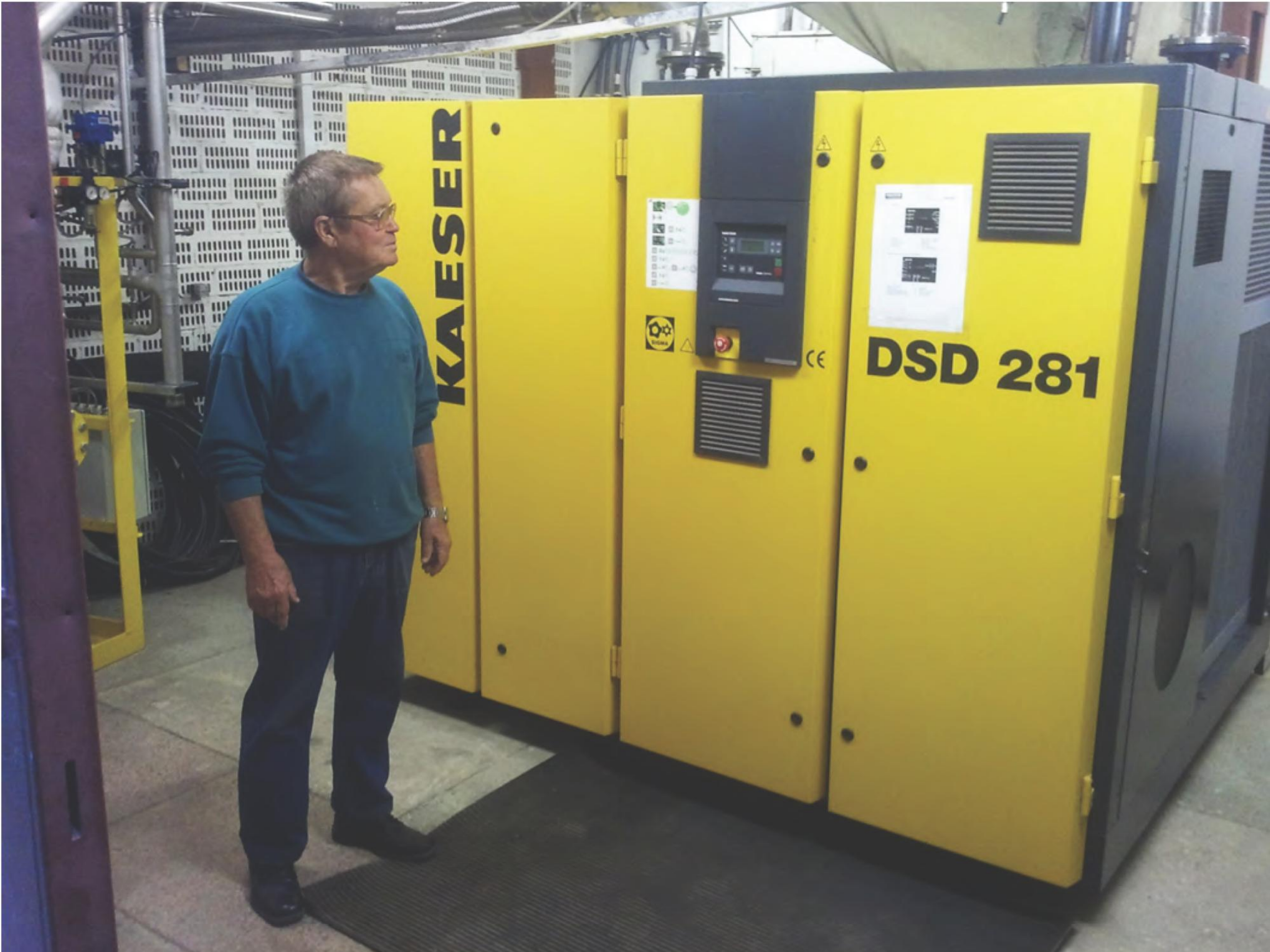}
\caption{The cryogenic system of the Troitsk-nu-mass experiment, planned
to be used in TASTE. }
\label{fig:cryo}
}
\end{figure}

\begin{figure}
\center{%
\includegraphics[width=\linewidth]{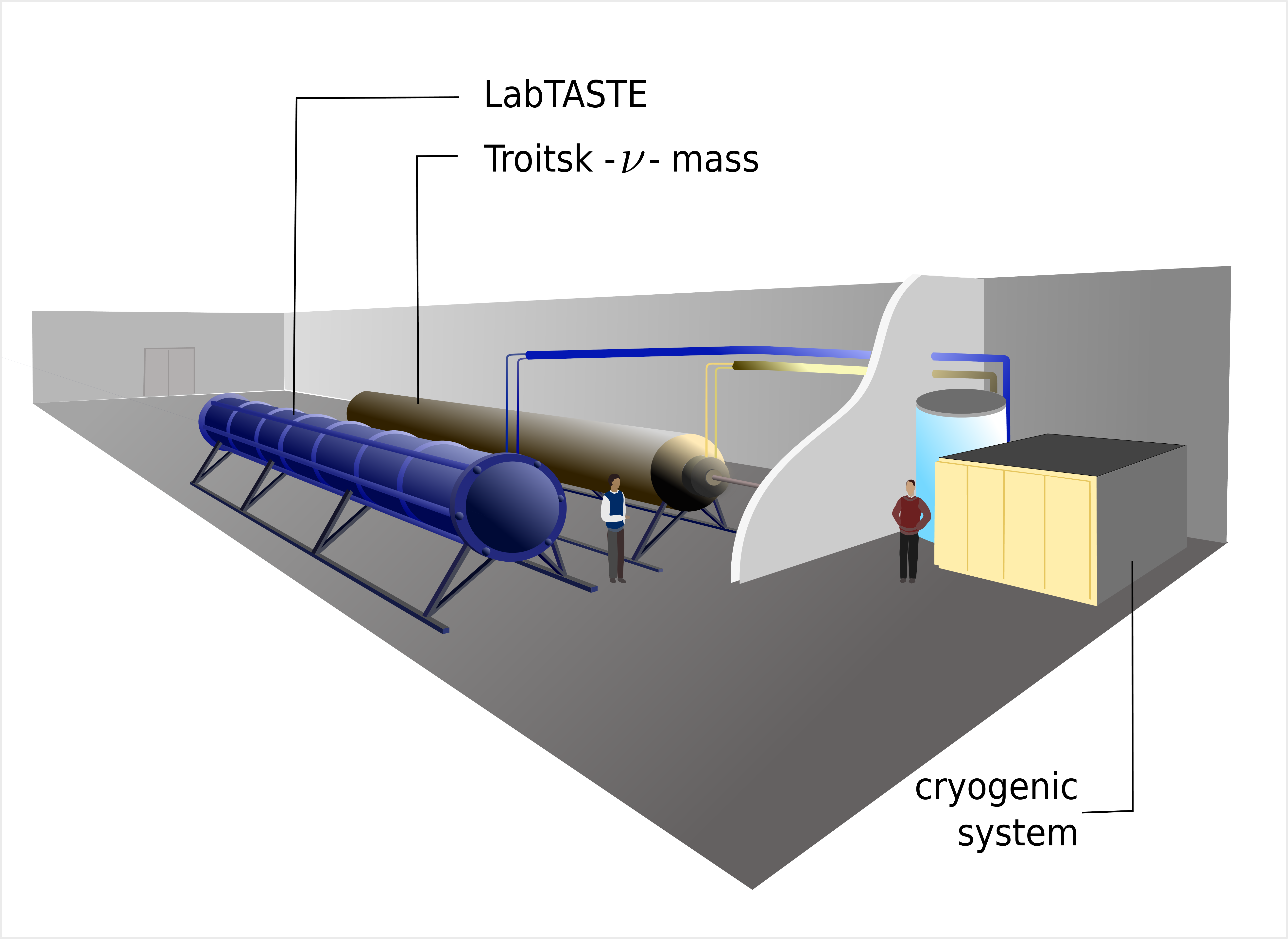}
\caption{
The artist's view of LabTASTE in the Troitsk-nu-mass experimental hall.
}
\label{fig:LabTASTE}
}
\end{figure}

In our proposal, we aim at the maximal usage of available resources and
plan to benefit from the cryogenic equipment of the Troitsk-nu-mass
experiment \cite{nu-mass} in INR. It includes a LINDE TCF-50 helium plant
with 60~l/h capability equipped with cryogenic transport and quench
protection systems, see photos in Fig.~\ref{fig:cryo}.
At the first stage of the project, LabTASTE, the system will be used in
turns with Troitsk-nu-mass, Fig.~\ref{fig:LabTASTE}.
When the full TASTE magnet is available,
the cryogenic system may be rearranged for the full-time use in TASTE.

\red After the main data taking program with vacuum bores, the stage of
the experiment with gas filling is anticipated to improve the sensitivity
at higher axion masses $m_{a}$, see Sec.~\ref{sec:helio:Pconv}.
We plan to use $^{4}$He or $^{3}$He for filling. The number of pressure
steps necessary for scanning the required mass range may be estimated
based on the CAST experience and discussion in Ref.~\cite{9-12}, see
Eqs.~(9) -- (12) there. For $L=12$~m and $m_{a} \lesssim 0.1$~eV, the
relative width of the mass range at the resonance is $\Delta m_{a}/m_{a}
\sim 3\%$, which transforms into $\sim 6\%$ variation in density. We will
need $\sim 30$ density settings, each of one day data taking, to cover
$m_{a} \lesssim 0.1$~eV. Moving to higher masses is more tricky, and our
estimate is to have $\sim 300$ density settings to reach $m_{a} \sim
0.5$~eV at the resonant sensitivity. More details of the gas phase will be
determined at the stage of the technical design of the magnet. \black

\section{X-ray photon collection}
\label{sec:X}
\subsection{X-ray optics}
\label{sec:X:opt}
\begin{figure}
\center{%
\includegraphics[width=0.75\linewidth]{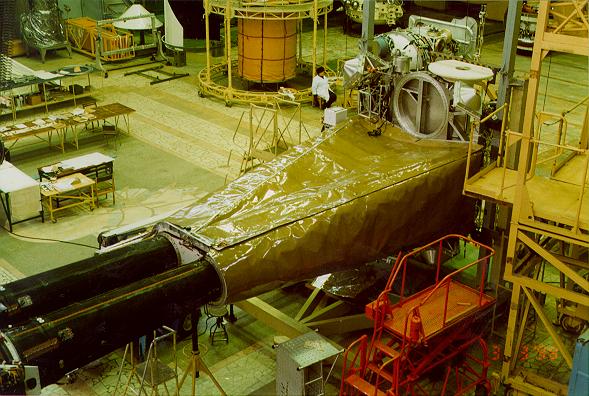}
\caption{Covered part of the SODART optical block with two telescopes.
}
\label{fig:SODART}
}
\end{figure}

Though focusing of energetic X-ray photons is not an easy task, numerous
X-ray telescopes have been developed for space-based astronomical
instruments (a brief overview of their relevant parameters is given e.g.\
in Ref.~\cite{NGAH}). In 1990s, the Soviet--Danish Roentgen Telescope
(SODART)~\cite{SODART-mirror} has been developed and manufactured for the
Spectrum-Roentgen-Gamma (SRG) space observatory which, however, has never
been launched. The modern version of the SRG satellite, being considered
for launch in 2018, will carry other scientific instruments. We propose to
use one of two SODART X-ray mirrors, Fig.~\ref{fig:SODART},
in the TASTE project.

\begin{figure}
\center{%
\includegraphics[width=0.75\linewidth]{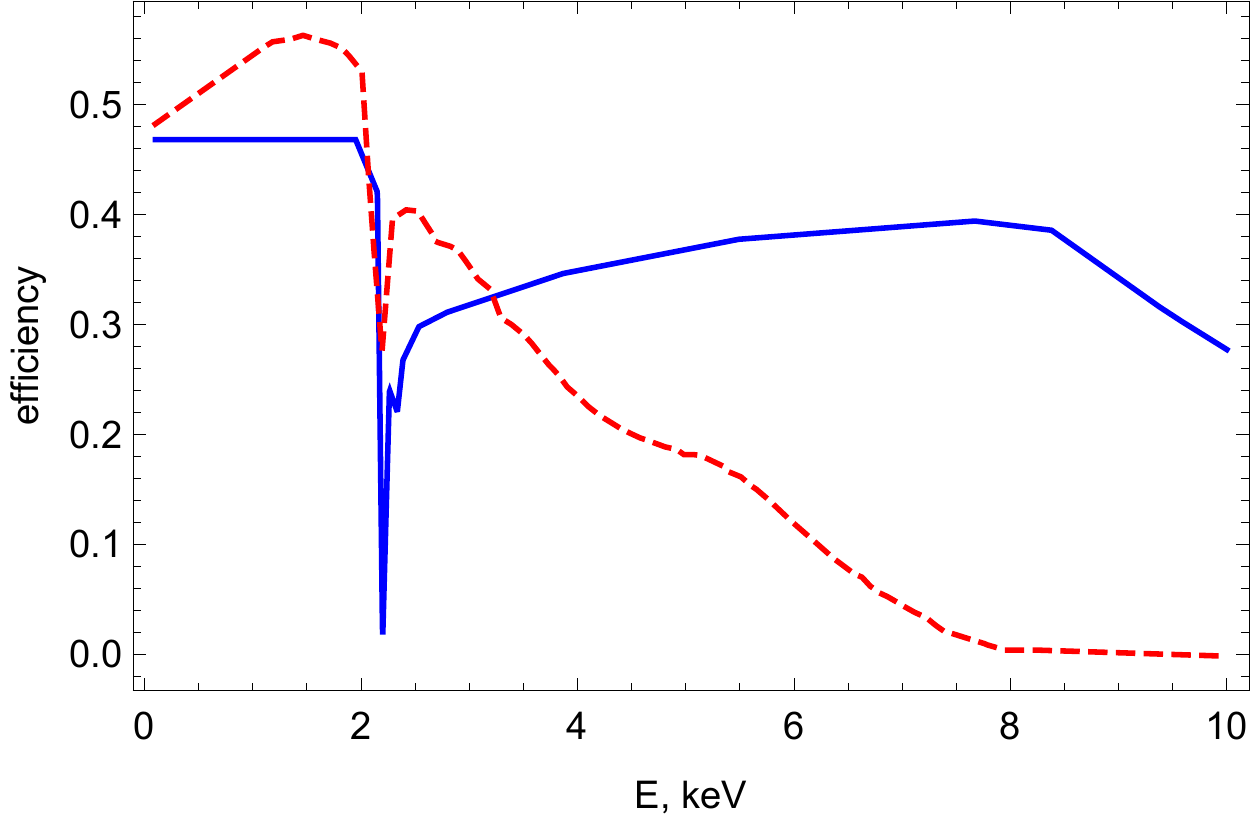}
\caption{\red The energy dependence of the X-ray optics efficiency for
SODART (full line, Ref.~\cite{SODART-preprint}) and CAST (dashed line,
Ref.~\cite{CASTeff}).}
\label{fig:SODARTeff}
}
\end{figure}

The SODART mirror \cite{SODART-mirror} consists of 143 nested aluminium
foil shells divided into quadrant and forming the cone which approximates
the Wolter~I geometry. The thickness of the shells is 0.4~mm, the length
of each one is 20~cm. The reflecting surface was prepared with a lacquer
coating technique. The telescope's diameter is 60~cm, the focal length is
8~m and the mass of the mirror is $\sim 100$~kg. The working energy range
of the instrument is $\sim (0.2-20)$~keV. The telescope has been tested
extensively in laboratory conditions \cite{SODART-testing, Bart} and its
performance has been simulated \cite{SODART-sim, SODART-Springer,
SODART-preprint}. Figure~\ref{fig:SODARTeff}
presents the
energy dependence of the on-axis effective area in the units of the ideal
geometric one~\cite{SODART-preprint}. For comparison, we plot also the
similar quantity for CAST \cite{CASTeff}. Weighted by the effective
sectrum of the Sikivie solar axions and integrated over energy, this gives
the values of the optics efficiency $\epsilon_{0}$, given in
Table~\ref{tab:FoM}. Note that preliminary simulations
indicated~\cite{SODART-Springer} flat efficiency within (0.1 -- 1)~keV
which may be important for searches of solar axions interacting with
electrons. \black

These SODART values
are used in our baseline design; if the telescope is not available, other
options will be considered; \red if it is, more detailed simulations and
measurements will be performed. \black

\subsection{X-ray photon detector}
\label{sec:X:det}
For the TASTE experiment, we plan to select the appropriate X-ray detector
through additional RnD studies. Options to be considered include several
solid-state detectors under development for astrophysics, high-energy
physics and axion searches. The approaches followed by various groups
participating in TASTE are described e.g.\ in Refs.~\cite{Debrin, det-FTI,
det-IKI}.

One of the most challenging parameters of the detector is its low
background. While present-day background rate values for astrophysical
detectors are too high for our purposes, they are dominated by cosmic-ray
contamination, which will be reduced by a combination of the passive
shielding and a dedicated veto system. The detectors themselves will be
tested and the shielding will be designed in the Low-background
Measurement Laboratory in Baksan Neutrino Observatory of INR.
\red Several approaches will be combined in the X-ray detecting unit to
fulfill the condition of extremely low dark counting rate, including:

-- reduction of the detector electronic noise by the detector cooling;

-- reduction of the front-end electronics noise by its cooling;

-- increase of the signal amplitude via avalanche multiplication in the
detector;

-- unification of the detector response and the use of its features for
rejection of noise signals.

To realize the approaches, a new construction of a silicon detector will
be developed and implemented in the DAQ system with digital processing of
the signals.  \black
Details of
the detector design will be discussed elsewhere.

\section{Tracking and infrastructure}
\label{sec:platform}
As we discussed above, our goal is to track the Sun for $\epsilon
_{t}=0.5$ fraction of the time throughout the year. This determines
certain requirements for the movable mount of the helioscope.

\begin{figure}
\center{%
\includegraphics[width=0.75\linewidth]{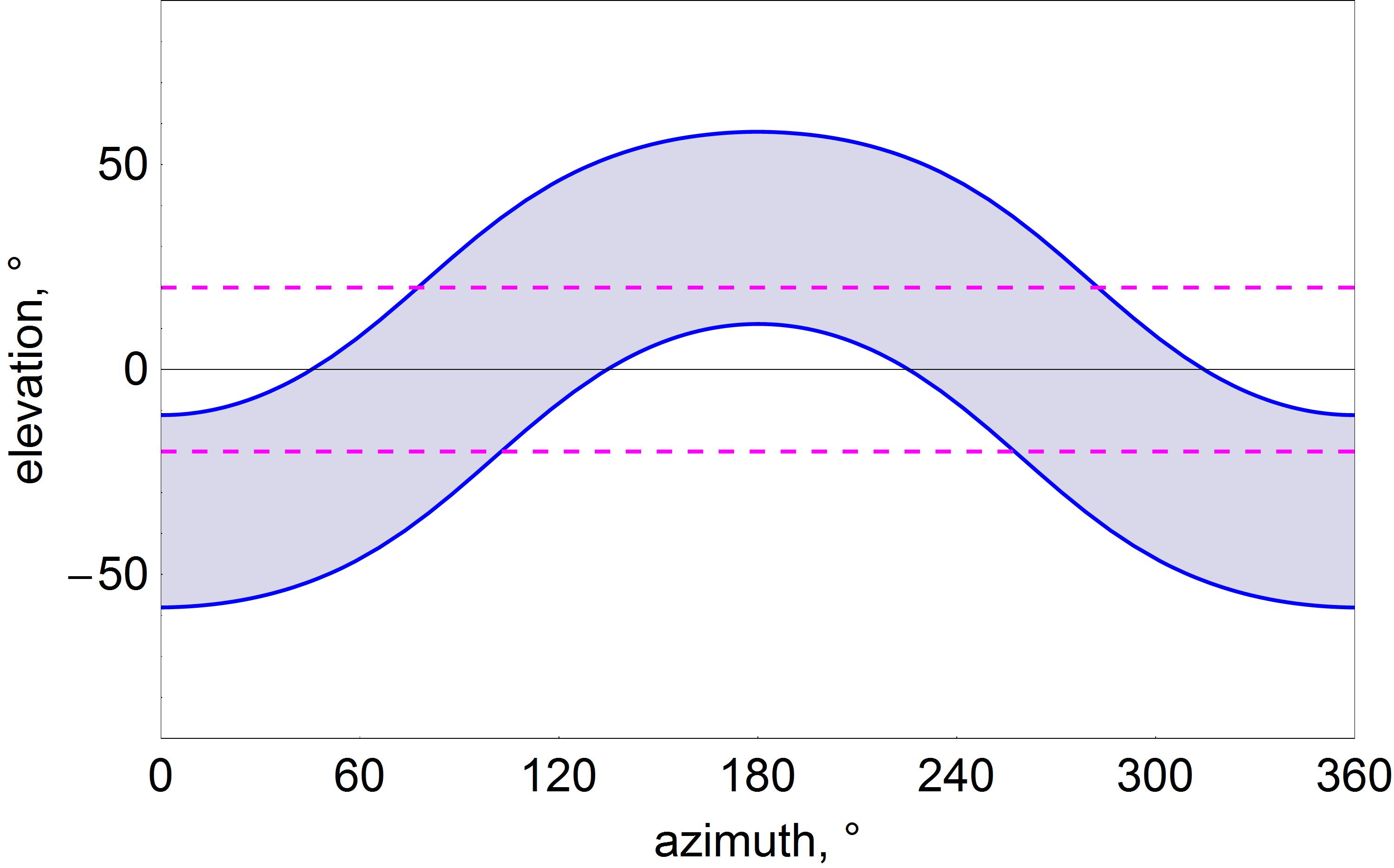}
\caption{Part of the sky spanned by the Sun throughout the year, in
horizontal coordinates (azimuth and elevation) for the latitude of
Troitsk. Dashed horizontal lines bound the range $\pm 20^{\circ}$
corresponding to $\epsilon _{t}\approx0.5$. }
\label{fig:tracking}
}
\end{figure}

In Fig.~\ref{fig:tracking},
we present the part of the sky spanned by the Sun throughout a year, for
the geographical latitude of Troitsk, $55.5^{\circ}$ North. A
straightforward calculation demonstrates that $\epsilon _{t}=0.5$ is
reached if the helioscope can move by $\pm 20^{\circ}$ above and below the
horizon, while rotating by 360$^{\circ}$ in the horizontal plane. It is
not beneficial to constrain the azimuthal movement at the price of
increased inclination range.

We estimate the total mass of the helioscope tube as $\sim 12$~tons. This
number demonstrates a serious advantage of the iron-free magnet design:
for comparison, the CAST magnet (having $\sim 100$ times smaller aperture
area) weights 27.5~tons. In addition, we estimate the mass of the X-ray
telescope support as $\sim 1$~ton (the telescope itself, designed for
launching in space, weights less than 0.1~ton), and another $\sim 1$~ton
for the detector shielding. For the total mass of $\sim 14$~tons for the
moving part of the installation, it is a nontrivial task to find or
construct the mount; however, solutions exist in industry, gamma-ray and
radio astronomy. We will determine the exact design of the moving mount at
subsequent stages of the project realization.

An important part of the helioscope is the vacuum vessel which hosts the
magnet and the cryostat. For the one-section magnet, LabTASTE, we plan to
use the vessel of the decommissionned old Troitsk-nu-mass spectrometer,
which is 7~m long and has a diameter of 1.2~m.

\begin{figure}
\center{%
\includegraphics[width=\linewidth]{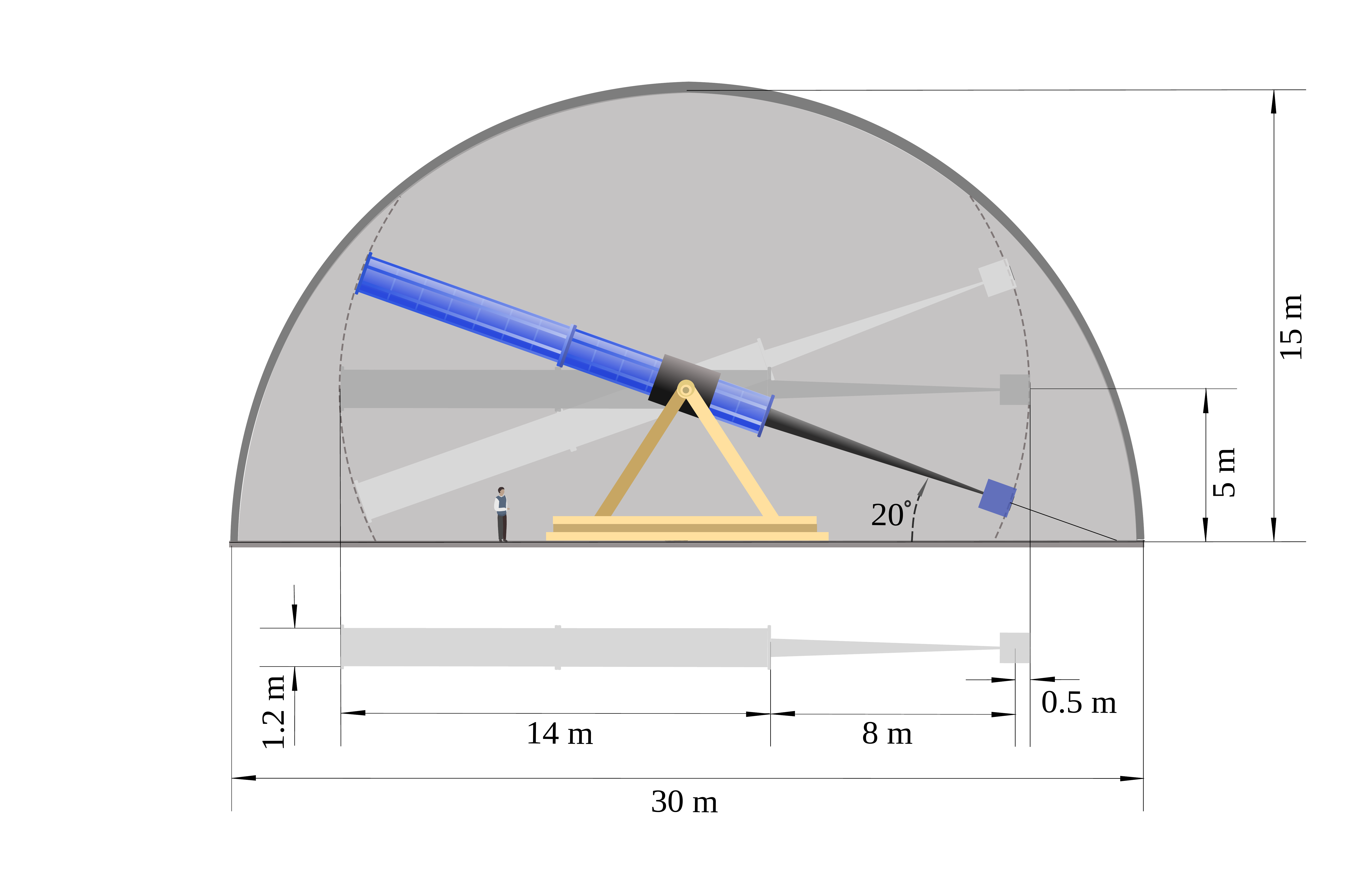}
\caption{Geometry of the TASTE helioscope in a half-cylinder warm hangar.
}
\label{fig:TASTEscheme}
}
\end{figure}

\begin{figure}
\center{%
\includegraphics[width=\linewidth]{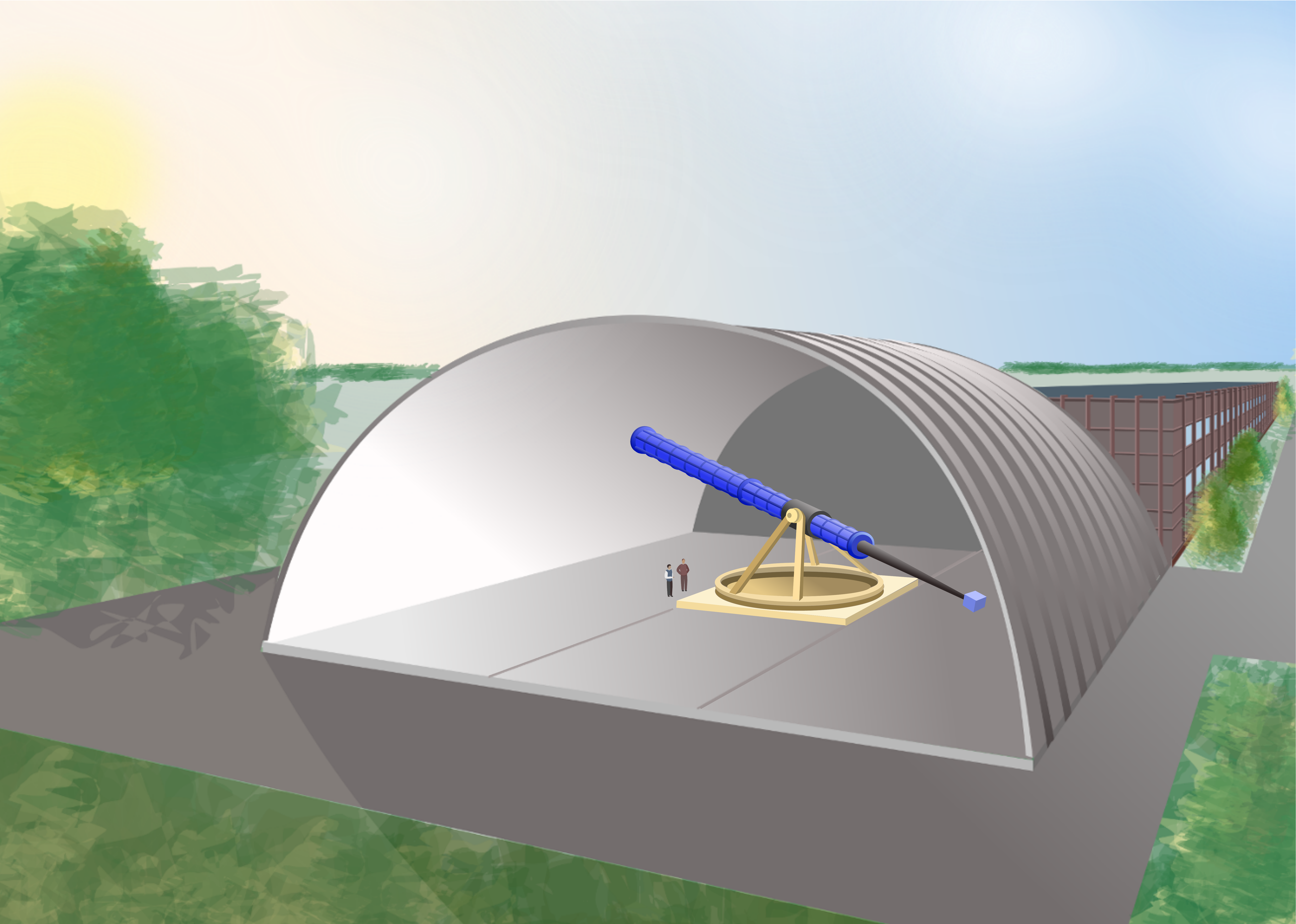}
\caption{The artist's view of the TASTE experiment in a warm hangar
attached to INR building~101 in Troitsk. The Troitsk-nu-mass cryogenic
system is located in this building. }
\label{fig:TASTEview}
}
\end{figure}

The proposed geometry of the helioscope is shown in
Fig.~\ref{fig:TASTEscheme}.
We plan to put it in a commercially available half-cylinder warm hangar
with the square base of 30~m$\times$30~m, attached to the INR building
where Troitsk-nu-mass is located (see Fig.~\ref{fig:TASTEview}
for the artist's view). The INR campus in Troitsk possesses all required
infrastructure, including high-power electric line built for the Moscow
Meson Factory (a linear proton accelerator in INR), laboratory space and
mechanical workshops.

\section{Discussion and conclusions}
\label{sec:concl}

\subsection{Costs and timeline}
\label{sec:concl:cost}
\begin{figure}
\center{%
\includegraphics[width=\linewidth]{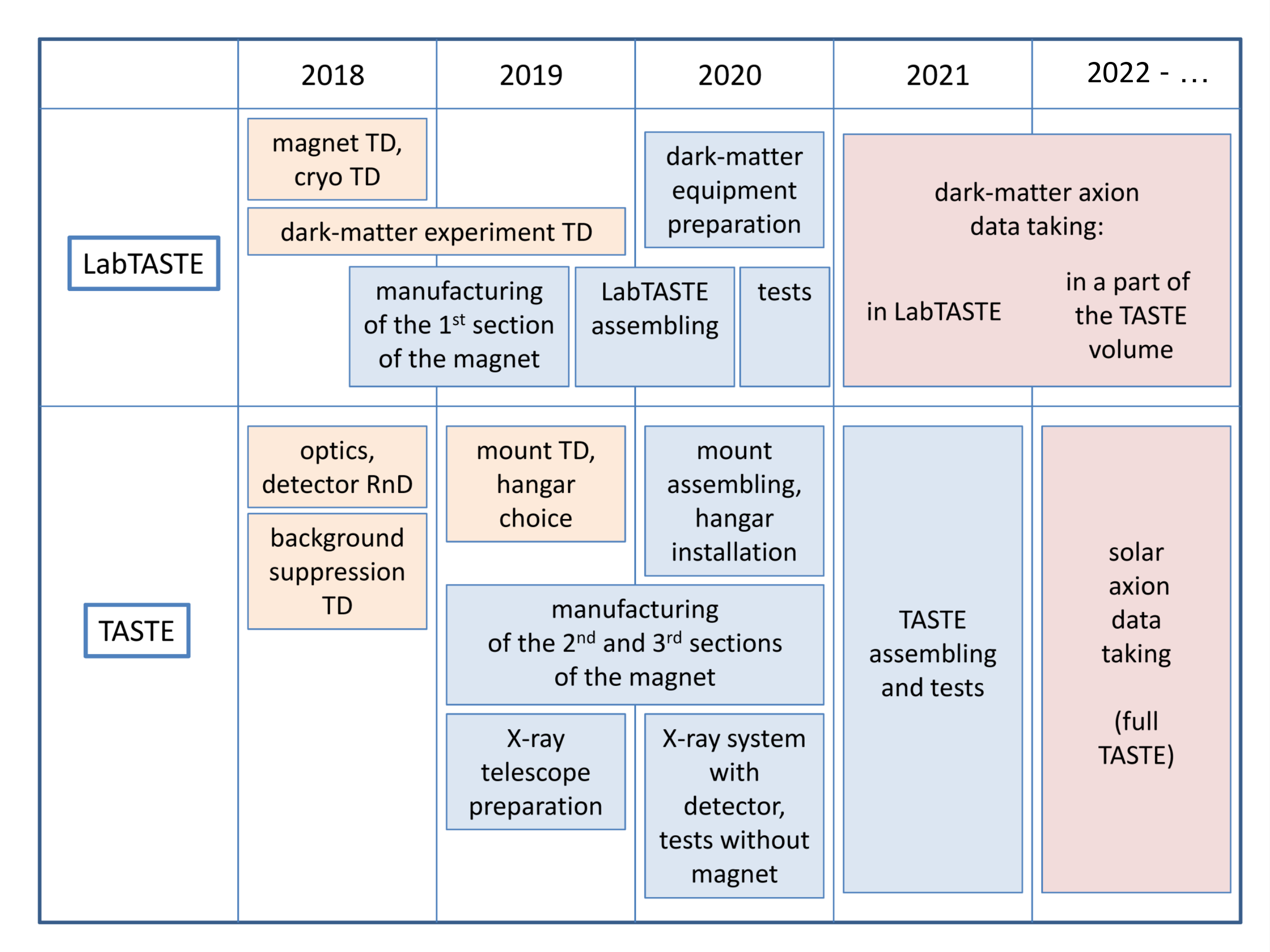}
\caption{TASTE estimated timeline.
}
\label{fig:timeline}
}
\end{figure}

\begin{figure}
\center{%
\includegraphics[width=\linewidth]{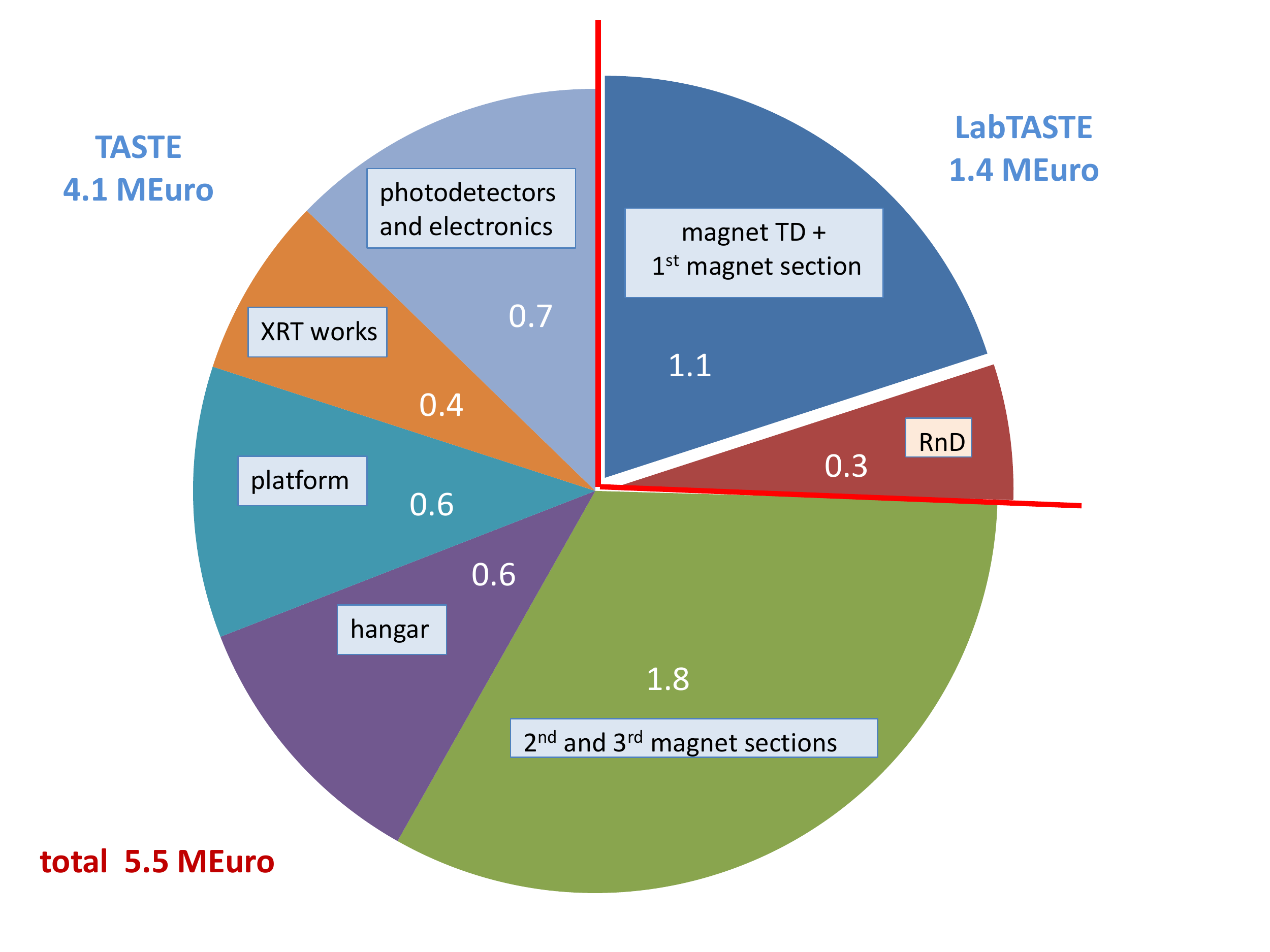}
\caption{TASTE estimated costs.
}
\label{fig:costs}
}
\end{figure}

As discussed above, we preview two stages of the experiment, LabTASTE and
the full TASTE. Subject to available funds,
these projects may be realized in parallel, and this is the scenario we
imply in the timeline presented in Fig.~\ref{fig:timeline}.
\red Our preliminary plan includes 3 years of data taking under the vacuum
conditions, then installation of the gas filling system and two
additional gas runs: 1 month to explore $m_{a} \lesssim 0.1$~eV and 1
year to reach $m_{a}\sim 0.5$~eV. \black  A rough estimate of the
budget is given in Fig.~\ref{fig:costs}.
This estimate, of course, does not include materials and
equipment already available: the superconducting cable, the helium
plant, the vacuum vessels and the X-ray telescope, as well as available
infrastructure at the INR campus in Troitsk.

\subsection{Scientific opportunities beyond solar $\g$}
\label{sec:concl:opport}

\subsubsection{Dark-matter axion searches}
\label{sec:concl:opport:DM}
Dark matter axion searches can be conducted during LabTASTE stage and
continued inside the full TASTE magnet. Rather large diameter, 60~cm, of
TASTE magnet is beneficial here. First, it can be used for RnD studies
for MADMAX, see Sec.~\ref{sec:motiv:experiments:DM}. TASTE bore area is
3.5 times smaller, and, additionally, magnetic field strength is a factor
of 2.8 smaller than required by the MADMAX design. Therefore, TASTE will
be a factor of 15 short for an average dark matter axion signal searches
using that concept. However, TASTE can be used for high amplitude
transient signals searches using broadband MADMAX-like design or even
simply employing dish antenna strategy \cite{Horns:2012jf}.

In models where Peccei-Quinn phase transition occurs after inflation, the
axion dark matter is fragmented  into miniclusters of large density
\cite{Kolb:1993hw}. During encounters with miniclustres the expected
signal in axion detectors would increase by a factor of $10^8$, however,
such encounters are rare, they occure on average once in $10^5$ years. On
the other hand, small fraction of miniclusters is destroyed in
gravitational collisions with stars. Density in resulting tidal streams is
smaller, but a probability of a crossing is larger. As a result
\cite{Tinyakov:2015cgg}, signal increase by an order of magnitude may be
expected every $\sim$20 years due to crossings of a  tidal stream
(smaller/larger amplitudes are more often/rare). Large amplitude signals
can also be caused by possible trapping of an axion minicluster in the
Solar system and by gravitational lensing of dark matter streams
\cite{Zioutas:2017klh}. In order not to miss the actually unpredictable
timing of such short duration signals (few days for a tidal streams from
miniclusres), a network of co-ordinated axion antennae is required,
preferentially distributed world-wide, and TASTE can be part of such
network. Prospects for TASTE here are even brighter for the ALP searches
with increased coupling to photons, covering yet unexplored territories in
their parameter space.

\subsubsection{Solar axions with the electron coupling}
\label{sec:concl:opport:electron}
In certain models, axions or ALPs interact with electrons by means of the
Yukawa term in the Lagrangian,
\[
-g_{ae}\,a\,\bar e \gamma_{5} e,
\]
where $e$ is the electron spinor field, $\gamma _{5}$ is the chiral Dirac
matrix and $g_{ae}$ is the dimensionless Yukawa coupling. The solar axion
flux due to this coupling has been calculated in Ref.~\cite{Redondo-e}; it
peaks at energies $\sim 1$~keV and is therefore detectable by a helioscope
provided it has a sufficiently low detection energy threshold. While the
emission rate is proportional to $g_{ae}^{2}$, the detection in a
helioscope is still driven by $\g^{2}$, and therefore this method
constrains the product $g_{ae}\g$ (unless the Sikivie flux dominates).
The limits of this kind have been obtained by CAST \cite{CAST-e}. They are
however weaker than the present-day astrophysical limits from red-giant
energy losses \cite{RG}. It is interesting to note that numerous studies
of white-dwarf cooling, based either on the luminosity function, e.g.\
Ref.~\cite{WD-LF}, or on long-term observations of pulsating white dwarfs,
e.g.\ Ref.~\cite{WD-osc1, WD-osc2}, continue to favour a nonzero value of
$g_{ae}\sim 10^{-13}$ (see Ref.~\cite{Maurizio} for a discussion). The
helioscope sensitivity to $\sqrt{g_{ae}\g}$ scales with the same figure of
merit $F^{1/4}$, Eq.~(\ref{Eq:FoM}); \red though TASTE will be hardly able
to probe $g_{ae}\sim 10^{-13}$, the most interesting scenario \black where
white-dwarf cooling and gamma-ray transparency are both explained by the
effects of one and the same particle \red will be probed through $\g$.
\black

\subsubsection{Other light particles}
\label{sec:concl:opport:DE}
\textit{Chameleons} are hypothetical scalar particles proposed
\cite{chame1, chame2} to explain the accelerated expansion of the Universe
(``dark energy''). Their interactions with matter result in the appearence
of the effective particle mass which depends on the ambient energy
density. In certain cases, they possess the interaction (\ref{*}) and
represent, therefore, a subclass of ALPs. However, the density dependence
of the mass makes their phenomenology quite different from the general ALP
case. They can be created in magnetic fields inside the Sun and detected
in a helioscope in a way similar to usual ALPs \cite{chame-helioscope}.
This has been used to constrain chameleon parameters with CAST
\cite{chame-CAST}. These constraints may be improved with TASTE provided
its detector is sensitive to sub-keV photons.

\textit{Paraphotons} \cite{Okun}, or hidden photons, are hypothetical
vector bosons of an additional $U(1)$ gauge group not present in the
Standard Model. Abelian gauge bosons mix in their kinetic terms
\cite{Holdom, Dienes}, and this makes the photon/paraphotons conversion
possible. Helioscopes can be used to constrain parameters of these
hypothetical vector bosons, see e.g.\ Refs.~\cite{Redondo-para, ST-para,
Redondo-atlas}. Since the external magnetic field is not required for the
conversion, switching off the magnet while continuing solar tracking might
help to distinguish paraphotons from ALPs in the case of positive
detection.

\subsection{Brief conclusions}
\label{sec:concl:concl}
To summarize, we propose a multi-purpose discovery experiment to search
for axions and other hypothetical light particles, predicted by extensions
of the Standard Model of particle physics and motivated by recent
astrophysical observations. Our projected device, with its total cost on
the scale of $\sim 5 $~MEuro, would test, on the timescale of less than
5~years, several models of the anomalous transparency of the Universe,
dark matter and even dark energy, as well as a particular part of the
parameter space relevant for the axion solution of the strong CP problem.
The results of the project would shed light on the unexplored processes
involving light particles in hot plasmas where thermonuclear reactions
take place.

\acknowledgments
We thank our colleagues from the worldwide axion-searching community,
notably Maurizio Giannotti, Igor Irastorza, Axel Lindner,
Javier Redondo, Andreas Ringwald and Yannis Semertzidis, for numerous
inspiring discussions; Valery Rubakov and the INR
administration for encouragement and helpful interest in the project; Oleg
Kazachenko, Vladimir Kekelidze, Vyacheslav Klyukhin, Sergey Kozub and
Nikolay Mezentsev for helpful discussions related to the magnet; Alexander
Blinov for consultations related to the moving platform.


\end{document}